\documentclass[aps,prd,longbibliography,reprint,superscriptaddress,preprintnumbers,floatfix,nofootinbib,notitlepage,showkeys]{revtex4-1}

\usepackage{amsmath,amssymb,amsfonts,mathtools} 
\usepackage{hepunits}
\usepackage[normalem]{ulem}
\usepackage[utf8]{inputenc} 
\usepackage[T1]{fontenc}
\usepackage{hyperref}
\usepackage{graphicx}
\usepackage{latexsym}
\usepackage{bbm}
\usepackage{epstopdf}
\usepackage{color}
\usepackage{braket}
\usepackage{hyperref}
\hypersetup{colorlinks=true,citecolor=OliveGreen,linkcolor=BrickRed,urlcolor=MidnightBlue}
\usepackage{lineno}
\usepackage{slashed}
\usepackage{placeins} 
\usepackage{diagbox}
\usepackage{dcolumn}
\usepackage{xfrac}
\usepackage[dvipsnames]{xcolor}

\newcommand{\dd}[1]{\mathop{\mathrm{d}#1}}

\newcommand{\Ns}{N_{\rm s}}
\newcommand{\Nt}{N_{\rm t}}
\newcommand{\Tc}{T_{\rm c}}
\newcommand{\CF}{C_{\rm F}}

\newcommand{\GE}{G_\mathrm{E}}
\newcommand{\GB}{G_\mathrm{B}}
\newcommand{\GEB}{G_{\mathrm{E},\mathrm{B}}}
\newcommand{\ZE}{Z_\mathrm{E}}
\newcommand{\kappaE}{\kappa_\mathrm{E}}
\newcommand{\kappaB}{\kappa_\mathrm{B}}
\newcommand{\kappaEB}{\kappa_{\mathrm{E},\mathrm{B}}}
\newcommand{\tauT}{\tau T}
\newcommand{\tauf}{\tau_\mathrm{F}}
\newcommand{\tauF}{\tau_\mathrm{F}} 

\newcommand{\rhoEB}{\rho_{\mathrm{E},\mathrm{B}}}
\newcommand{\rhoE}{\rho_{\mathrm{E}}}
\newcommand{\rhoB}{\rho_{\mathrm{B}}}

\newcommand{\ReTr}{\textrm{Re}\,\textrm{Tr}\,}
\newcommand{\MSb}{\overline{\textrm{MS}}}

\newcommand{\be}{\begin{equation}} 
\newcommand{\ee}{\end{equation}}
\def\ba#1\ea{\begin{align}#1\end{align}}
\newcommand{\bea}{\begin{eqnarray}} 
\newcommand{\eea}{\end{eqnarray}}
\def\lsim{\mathrel{\raise.3ex\hbox{$<$\kern-.75em\lower1ex\hbox{$\sim$}}}}
\def\gsim{\mathrel{\raise.3ex\hbox{$>$\kern-.75em\lower1ex\hbox{$\sim$}}}}

\renewcommand{\sfrac}[2]{\ensuremath{#1/#2}}
\newcommand{\TM}{{\sfrac{T}{M}}~}

\begin{document}
\title{Heavy quark diffusion coefficient with gradient flow}
\author{Nora Brambilla}
\email{nora.brambilla@ph.tum.de}
\affiliation{Physik Department, Technische Universit\"at M\"unchen,
James-Franck-Strasse 1, 85748 Garching, Germany}
\affiliation{Institute for Advanced Study, Technische Universit\"at M\"unchen,
Lichtenbergstrasse 2a, 85748 Garching, Germany}
\affiliation{Munich Data Science Institute, Technische Universit\"at M\"unchen, \\
Walther-von-Dyck-Strasse 10, 85748 Garching, Germany}

\author{Viljami Leino}
\email{viljami.leino@tum.de}
\affiliation{Physik Department, Technische Universit\"at M\"unchen,
James-Franck-Strasse 1, 85748 Garching, Germany}

\author{Julian Mayer-Steudte}
\email{julian.mayer-steudte@tum.de}
\affiliation{Physik Department, Technische Universit\"at M\"unchen,
James-Franck-Strasse 1, 85748 Garching, Germany}
\affiliation{Munich Data Science Institute, Technische Universit\"at M\"unchen, \\
Walther-von-Dyck-Strasse 10, 85748 Garching, Germany}

\author{Peter Petreczky}
\email{petreczk@bnl.gov}
\affiliation{Physics Department, Brookhaven National Laboratory,
  Upton, New York 11973, USA}


\collaboration{TUMQCD Collaboration}
\noaffiliation

\date{\today}
\preprint{TUM-EFT 166/21}

\begin{abstract}
We calculate chromoelectric and chromomagnetic correlators in quenched QCD at
$1.5T_c$ and $10^4 T_c$, with the aim to estimate the heavy quark diffusion coefficient
at leading-order in the  inverse heavy quark mass expansion, $\kappaE$, as well as
the coefficient of the first mass-suppressed correction, $\kappaB$. We use gradient flow
for noise reduction. At $1.5T_c$ we obtain $1.70 \le \sfrac{\kappaE}{T^3} \le 3.12$
and $1.03< \sfrac{\kappaB}{T^3} < 2.61$. The latter implies that the mass-suppressed 
effects in the heavy quark diffusion coefficient are 20\% for bottom quarks and 34\% for charm quarks at this temperature. 
\end{abstract}

\maketitle

\section{Introduction}\label{sec:intro}
The behavior of a heavy quark moving in a strongly coupled quark gluon plasma (sQGP) can be described by a set of transport coefficients.
In particular, the equilibration time of the heavy quarks is related to the heavy quark diffusion.
This diffusion can be described as a Brownian motion and, hence, by a Langevin equation that depends on three
related transport coefficients~\cite{Moore:2004tg}: the heavy quark momentum diffusion coefficient $\kappa$, 
the heavy quark diffusion coefficient $D_\mathrm{s}$, and the drag coefficient $\eta$. In thermal equilibrium these coefficients are related
as $D_\mathrm{s}  = \sfrac{2T^2}{\kappa}$ and  $\eta=\sfrac{\kappa}{(2MT)}$, with
$M$ being the heavy quark mass.
The heavy quark momentum diffusion coefficient is known  in perturbation theory up to mass-dependent contributions at next-to-leading-order (NLO) accuracy~\cite{Moore:2004tg,Svetitsky:1987gq,Caron-Huot:2008dyw}.
We will label this leading term in the \TM expansion as $\kappaE$.
Moreover, the first mass-dependent contribution when expanding the diffusion coefficient with respect to \TM has been studied in Refs.~\cite{Bouttefeux:2020ycy,Laine:2021uzs}, and will be labeled $\kappaB$. 
It is sensitive to chromomagnetic screening and, therefore,
is not calculable in perturbation theory~\cite{Bouttefeux:2020ycy}.
Apart from describing the equilibration time of the heavy quark in a plasma, the diffusion coefficient $\kappa$ is a crucial parameter entering the evolution equations which describe the out-of-equilibrium dynamics of heavy quarkonium in sQGP~\cite{Brambilla:2016wgg,Brambilla:2017zei,Brambilla:2019tpt}.

The NLO correction to $\kappaE$ is sizable~\cite{Caron-Huot:2008dyw}, thus calling into question the validity of the perturbative expansion, and inviting instead a strong coupling calculation.
Currently, the only analytical strong coupling calculations available are for supersymmetric Yang-Mills theories~\cite{Herzog:2006gh,Casalderrey-Solana:2006fio}, and
therefore nonperturbative lattice QCD calculations for the heavy quark diffusion coefficient are heavily desired. 
However, a direct calculation of the transport coefficients on the lattice can be very challenging, as it involves a reconstruction of the spectral functions from the appropriate Euclidean time correlation functions. The transport coefficient is then defined as the width of the transport peak, a narrow peak at low energy $\omega$. Reconstruction of the spectral function in the presence of a transport peak is a  challenging problem, especially since the width of this peak is inversely proportional to the heavy quark mass $M$~\cite{Petreczky:2005nh}. Moreover, the Euclidean time correlators are relatively   
insensitive to small widths~\cite{Aarts:2002cc,Petreczky:2005nh,Petreczky:2008px,Ding:2012sp,Borsanyi:2014vka,Ding:2021ise}.

The problem of the transport peak can be circumvented by the use of the effective field theory approach. 
In particular, the heavy quark momentum diffusion coefficient can be related to correlators of field-strength tensor components. 
The leading contribution in the \TM expansion $\kappaE$ is related to a correlator of two chromoelectric fields $E$~\cite{Caron-Huot:2009ncn,Brambilla:2016wgg}, 
and the \TM correction is related to a correlator of two chromomagnetic fields $B$~\cite{Bouttefeux:2020ycy}. 
The associated spectral functions $\rhoEB(\omega)$ corresponding to these correlators do not have a transport peak, and the heavy quark diffusion coefficients $\kappaEB$ are defined as their 
$\omega \rightarrow 0$ limit. 
Moreover, the small-$\omega$ behavior is smoothly
connected to the UV behavior of the spectral function~\cite{Caron-Huot:2009ncn}. 

The chromoelectric correlator has been calculated on the lattice within this approach in the SU(3) gauge theory in the deconfined phase, i.e., for purely gluonic plasma~\cite{Meyer:2010tt,Francis:2011gc,Banerjee:2011ra,Francis:2015daa,Brambilla:2020siz}, using the multilevel algorithm for noise reduction~\cite{Luscher:2001up}.
It has also been studied out-of-equilibrium using classical, real-time lattice simulations in Refs.~\cite{Boguslavski:2020mzh,Boguslavski:2020tqz}.
During the writing of this paper, the first measurement of the mass-suppressed effects was reported in Ref.~\cite{Banerjee:2022uge}, also utilizing
the multilevel algorithm for noise reduction.

Recently, there has been a lot of interest in using gradient flow~\cite{Narayanan:2006rf,Luscher:2009eq,Luscher:2010iy} for noise
reduction instead of the multilevel algorithm. 
The gradient flow algorithm is a smearing algorithm that automatically renormalizes any gauge-invariant observables~\cite{Luscher:2010we,Luscher:2011bx} for a sufficiently large level of smearing.
The heavy quark diffusion coefficient $\kappaE$ has been measured with gradient flow 
in Ref.~\cite{Altenkort:2020fgs}.
Also, preliminary measurements of the chromomagnetic correlator required for $\kappaB$ have been performed with gradient flow and presented in conference proceedings by two groups~\cite{Altenkort:2021ntw,Mayer-Steudte:2021hei}.

In this work, we study both the chromoelectric and chromomagnetic correlators on the lattice using the gradient flow algorithm and determine the diffusion coefficient components $\kappaE$ and $\kappaB$ from the respective reconstructed spectral functions. In Sec.~\ref{sec:correlator} we recall the theory behind the required Euclidean correlators and show the raw lattice measurements of these correlators together with their continuum limits. In Sec.~\ref{sec:GEkappa} we then invert the spectral function and provide ranges for $\kappaEB$. The results are summarized in Sec.~\ref{sec:conc}.
Preliminary versions of these results have been published in a recent conference proceedings~\cite{Mayer-Steudte:2021hei}.

\section{Chromoelectric and Chromomagnetic correlators}\label{sec:correlator}
\subsection{Theory background and lattice setup}\label{subsec:theorybackg}
Heavy quark effective theory (HQET) provides a method to calculate the heavy quark diffusion coefficient
in the heavy quark limit $M\gg\pi T$ by relating it to correlators in Euclidean time.
The leading-order contribution $\kappaE$ to the heavy quark momentum diffusion coefficient $\kappa$ 
has been expressed in terms of the chromoelectric correlator $\GE$ in Refs.~\cite{Caron-Huot:2009ncn,Casalderrey-Solana:2006fio}:
\be\label{eq:gelat}
\GE(\tau) = -\sum_{i=1}^{3} 
 \frac{\left\langle \ReTr\left[U(1/T,\tau)E_i(\tau,\mathbf{0})U(\tau,0)E_i(0,\mathbf{0})\right]\right\rangle}{3\left\langle\ReTr U(1/T,0)\right\rangle}\,,
\ee
where $T$ is the temperature, $U(\tau_1,\tau_2)$ is a Wilson line in the Euclidean time direction, 
and $E_i$ is the chromoelectric field, which is discretized on the lattice as~\cite{Caron-Huot:2009ncn}
\begin{linenomath}\begin{align}
    E_i(\tau,\mathbf{x})=U_i(\tau,\mathbf{x})U_4(\tau,\mathbf{x}+\hat{i})-U_4(\tau,\mathbf{x})U_i(\tau,\mathbf{x}+\hat{4})\,.
\end{align}\end{linenomath}
Recently, the first correction in $\mathbf{v}^2$ to $\kappa$, known as $\kappaB$, has been put in relation to the  chromomagnetic correlator $\GB$~\cite{Bouttefeux:2020ycy}:
\be\label{eq:gblat}
\GB(\tau) = \sum_{i=1}^{3} 
 \frac{\left\langle \ReTr\left[U(1/T,\tau)B_i(\tau,\mathbf{0})U(\tau,0)B_i(0,\mathbf{0})\right]\right\rangle}{3\left\langle\ReTr U(1/T,0)\right\rangle}\, ,
\ee
where $B_i$ is the chromomagnetic field, which is herein discretized as:
\ba
    B_i(\tau,\mathbf{x})=\epsilon _{ijk}U_j(\tau,\mathbf{x})U_k(\tau,\mathbf{x}+\hat{j})\,.
    \label{eq:B_field_loop}
\ea
For both chromoelectric and chromomagnetic fields, we follow the usual lattice convention and absorb the coupling into the field definition:
$E_i\equiv gE_i$ and $B_i\equiv gB_i$.

The Euclidean correlators $\GE$ and $\GB$ are related to the respective heavy quark momentum diffusion coefficient contributions $\kappaE$ and $\kappaB$
by first obtaining the spectral functions $\rhoEB(\omega,T)$,
\ba
\GEB(\tau) &= \int_0^\infty \frac{\dd{\omega}}{\pi}\rhoEB(\omega,T)K(\omega,\tauT)\,,\label{eq:gefromrho}\\
\intertext{where}\nonumber\\
K(\omega,\tauT) &= \frac{\cosh\left(\frac{\omega}{T}\left(\tauT-\frac{1}{2}\right)\right)}{\sinh\left(\frac{\omega}{2T}\right)}\,,
\ea
and then taking the zero-frequency limit:
\be
\kappaEB \equiv \lim_{\omega\rightarrow 0} \frac{2T\rhoEB(\omega,T)}{\omega}\label{eq:kappalimd}\,.
\ee
The spectral function $\rhoE(\omega,T)$ for the chromoelectric correlator $\GE$ does not depend on the renormalization in the $a\rightarrow 0$ limit, 
however, the respective spectral function $\rhoB(\omega,T)$ for the chromomagnetic correlator $\GB$ does~\cite{Bouttefeux:2020ycy}.
On the other hand, both $\kappaE$ and $\kappaB$ are physical observables and the $\omega \rightarrow 0$ limit of the respective spectral functions does not depend on the renormalization.
The two contributions $\kappaE$ and $\kappaB$ can then be combined to give the full expression for the heavy quark momentum diffusion coefficient $\kappa$~\cite{Bouttefeux:2020ycy}:
\be
\kappa = \kappaE + \frac{2}{3}\langle \mathbf{v}^2 \rangle \kappaB\,.
\ee

In order to perform the needed lattice calculations, we use the MILC Code~\cite{MILC} to generate a set of pure-gauge SU(3) configurations using the standard Wilson gauge action.
The configurations are generated with the heat-bath and over-relaxation algorithms, 
where each lattice configuration is separated by at least 120 sweeps, each consisting of 15-20 over-relaxation steps and 5-15 heat-bath steps.
We consider two temperatures: a low temperature $1.5\Tc$, and a high temperature $10^4\Tc$, with $\Tc$ being the deconfinement phase transition temperature.
The temperatures are set by relating them to the lattice coupling $\beta=\sfrac{6}{g_0^2}$, which determines the lattice spacing $a$ via the scale setting~\cite{Francis:2015lha}. 
This scale setting relates $\beta$ to a gradient flow parameter $t_0$ via a renormalization-group-inspired fit form, which is then further related
to the temperature with $\Tc \sqrt{t_0}=0.2489(14)$~\cite{Francis:2015lha}.
For this study, we use lattices with varying numbers of temporal sites, $\Nt=20,~24,~28,~\mathrm{and}~34$, and with corresponding spatial extents of $\Ns=48$,~48,~56, and~68 sites. 
Based on our previous study~\cite{Brambilla:2020siz}, we do not expect there to be a notable dependence on the spatial size of the lattice.

To measure the Euclidean correlators we rely on the gradient flow algorithm~\cite{Narayanan:2006rf,Luscher:2009eq,Luscher:2010iy}.
The Yang-Mills gradient flow evolves the gauge fields $A_\mu$ toward the minimum of the Yang-Mills gauge action along a flow time $\tauf$:
\ba
    \Dot{B}_\mu &= D_\nu G_{\nu\mu},\ B_{\mu|\tauF=0}=A_\mu \\
    G_{\mu\nu} &= \partial_\mu B_\nu - \partial_\nu B_\mu + \left[ B_\mu,B_\nu \right],\ D_\mu =\partial_\mu + \left[ B_\mu,. \right]\,.
\ea
These equations are an explicit representation of
\ba\label{eq:baseflow}
    \partial_{\tauF}B_\mu(\tauF,x) = -g_0^2\frac{\delta S_{\mathrm{YM}}[B]}{\delta B_\mu (\tauF,x)}\,.
\ea

Adapting these equations for a pure-gauge lattice theory with link variables gives us the differential equation
\ba
    \Dot{V}_{\tau _F}(x,\mu)&=-g_0^2 \{ \partial_{x,\mu}S_{\mathrm{Gauge}}(V_{\tauF})\} V_{\tauF}(x,\mu)\label{eq:lat_flow}\\
    V_{\tauF}(x,\mu)|_{\tauF=0}&=U_\mu(x)\,,
\ea
where $V_{\tauF}$ are the flowed link variables. We choose $S_{\mathrm{Gauge}}$ to be the Symanzik action. 
The lattice simulations with $\Nt=20$, 24, and 28 are evaluated numerically with a fixed step-size integration scheme~\cite{Luscher:2010iy}, while the $\Nt=34$ lattice is evaluated with an adaptive step-size implementation~\cite{Fritzsch:2013je,Bazavov:2021pik}.
For further analysis, we need the data points from all lattices at the same flow time positions. 
Therefore, we use cubic spline interpolations with simple natural boundary conditions in order to provide the data along a common flow-time axis.
The full list of parameters and statistics are given in Table~\ref{tab:simulation_parameters}.
\begin{table}
\caption{Simulation parameters for the lattices.}
\label{tab:simulation_parameters}
    \centering
    \begin{ruledtabular}
    \begin{tabular}{cccdc}
        $T/\Tc$ & $N_\mathrm{t}$ & $N_\mathrm{s}$ & \multicolumn{1}{c}{$\beta$} & $N_\mathrm{conf}$\\\hline
        1.5 & 16 & 48 & 6.872 & 990 \\
            & 20 & 48 & 7.044 & 4290 \\
            & 24 & 48 & 7.192 & 4346 \\
            & 28 & 56 & 7.321 & 5348 \\
            & 34 & 68 & 7.483 & 3540 \\
      10 000 & 16 & 48 & 14.443 & 990 \\
            & 20 & 48 & 14.635 & 1890 \\
            & 24 & 48 & 14.792 & 2280 \\
            & 28 & 56 & 14.925 & 2190 \\
            & 34 & 68 & 15.093 & 1830
    \end{tabular}
    \end{ruledtabular}
\end{table}

The gradient flow evolves the unflowed gauge fields $A_\mu$ to the flowed fields $B_\mu$, which have been smeared with a flow radius $\sqrt{8\tauf}$.
This smearing systematically cools off the UV physics and automatically renormalizes the gauge-invariant observables~\cite{Luscher:2011bx}. 
This renormalization property of the gradient flow is especially useful for the correlators $\GEB$, which otherwise require renormalization on the lattice. The renormalization of the correlators can be calculated in lattice perturbation theory, like in Ref.~\cite{Christensen:2016wdo}, but the lattice perturbation theory may have poor convergence~\cite{Lepage:1992xa}.
In previous multilevel studies of the chromoelectric correlator~\cite{Francis:2015daa,Brambilla:2020siz},
a perturbative one-loop result for the chromoelectric field renormalization $\ZE$ was used~\cite{Christensen:2016wdo}.
As gradient flow automatically renormalizes gauge-invariant observables, such a factor $\ZE$ is not needed in this study, as has been observed already in the previous studies of $\GE$ using gradient flow~\cite{Altenkort:2020fgs}.
The continuum- and flow-time-extrapolated result for
$\GE$ at $1.5\Tc$ obtained using gradient flow agrees with the continuum extrapolated
results obtained using the multilevel algorithm and with the one-loop result for $\ZE$ at
the level of a few percent, indicating that for the $\beta$ range considered in the
calculations of $\GE$, the perturbative renormalization is fairly accurate. 
Moreover, in a recent lattice study of a different but similar operator, where a chromoelectric field was inserted into a Wilson loop~\cite{Leino:2021vop},
it was shown explicitly that $\ZE\rightarrow 1$ at sufficiently large flow times.
For chromomagnetic fields, renormalization is required both on the lattice and in continuum~\cite{Laine:2021uzs}. 
As the renormalization property of gradient flow is generic to all gauge-invariant observables~\cite{Luscher:2011bx}, the chromomagnetic correlator
should require no additional renormalization on the lattice either. 

On the other hand, since the gradient flow introduces a new length scale $\sqrt{8\tauf}$, 
we have to make sure it does not contaminate the measurements at the length scale of interest $\tau$ -- the separation between the field-strength tensor components.
The most basic condition for ensuring that the flow has enough time to smooth the UV regime, while preserving the physics at the scale $\tau$, would be
$a\lsim \sqrt{8\tauf} \lsim \sfrac{\tau}{2}$. The upper limit of this condition was further restricted in Ref.~\cite{Altenkort:2020fgs}, 
by inspecting the LO perturbative behavior of the flow~\cite{Eller:2018yje}, to be $\sfrac{(\tau-a)}{3}$. In our experience, slightly larger flow times are still fine; 
hence, we use a slightly relaxed limit:
\be\label{eq:trad_tauflims}
a\le \sqrt{8\tauf} \le \frac{\tau}{3}\,.
\ee
Moreover, we note that instead of dealing with the scales $\sqrt{8\tauf}$ and $\tau$ separately, the relevant scale for these Euclidean correlators
is in fact the ratio of the scales, $\sfrac{\sqrt{8\tauf}}{\tau}$. This can be inferred from the leading-order result of the chromoelectric correlator at finite flow time~\cite{Eller:2018yje}:
\be
\langle E(\tau,\tauf)E(0,\tauf)\rangle = \frac{g^2\delta^{ab}}{\pi^2}\sum_{n\in\mathbb{Z}} 
\frac{\delta_{ij}}{x_n^4}\left[(\xi^4_n + \xi^2_n +1) e^{-\xi_n^2}-1\right]\,,
\ee
where $\xi^2_n = x_n^2T^2/\tauf$ and $x_n = \tau + n/T$. Likewise, the early NLO result from Ref.~\cite{Eller:2021qpp} also shows an affinity to this ratio.
Using the units of $\sfrac{\sqrt{8\tauf}}{\tau}$, the condition of suitable flow times from Eq.~\eqref{eq:trad_tauflims} becomes
\be
\frac{a}{\tau} \le \frac{\sqrt{8\tauf}}{\tau} \le \frac{1}{3}\,.
\ee
We use these limits for both $\GE$ and $\GB$.

In order to reduce the discretization errors further, we define a tree-level improvement by matching the LO continuum perturbation theory result~\cite{Caron-Huot:2009ncn},
\be\label{eq:gepert}
\frac{\GE^\mathrm{LO}(\tau)}{g^2\CF} \equiv G^{\rm norm}(\tau) = \pi^2 T^4 \left[\frac{\cos^2(\pi\tauT)}{\sin^4(\pi \tauT)}+\frac{1}{3\sin^2(\pi\tauT)}\right]\,,
\ee
to the LO lattice perturbation theory result~\cite{Francis:2011gc},
\ba\label{eq:lolatpert}
\frac{\GE^\mathrm{LO,lat}(\tau)}{g^2\CF} &= 
\int_{-\pi}^{\pi} \frac{\mathop{\mathrm{d}^3 q}}{(2\pi)^3}
\frac{\Tilde{q}^2e^{\Bar{q}\Nt(1-\tauT)} + \Tilde{q}^2e^{\Bar{q}\Nt \tauT}}{3a^4\left(e^{\Bar{q}\Nt}-1\right)\sinh(\Bar{q})}\,,\\
\intertext{where}
\Bar{q} &= 2\mathrm{arsinh}\left(\frac{\sqrt{\Tilde{q}^2}}{2}\right)\,,\label{eq:barq} \\
\Tilde{q}^n &= \sum_{i=1}^3 2^n\sin^n\left(\frac{q_i}{2}\right)\,.\label{eq:tildeq}
\ea
We then define a tree-level improvement of $\GE$ as~\cite{Altenkort:2020fgs}
\be
\GE^\mathrm{imp}(\tauf,\tauT) = \frac{\GE^\mathrm{LO}(0,\tauT)}{\GE^\mathrm{LO,lat}(0,\tau)} \GE^\mathrm{measured}(\tauf,\tauT)\,,
\ee
where the improvement is restricted to zero-flow-time discretization effects, because the lattice perturbation theory result for Symanzik flow is not known.
We use the same tree-level improvement for the chromomagnetic correlator as for the chromoelectric correlator, since in the continuum limit these
correlators are the same at leading-order. We also used the clover discretization for the chromomagnetic correlators in addition to the one 
given in Eq.~\eqref{eq:B_field_loop}. The clover discretization was used in Ref.~\cite{Banerjee:2022uge}. We check that in the continuum
limit, the clover discretization and the one given by Eq.~\eqref{eq:B_field_loop} yield identical results within errors. The corresponding
analysis is discussed in Appendix~\ref{app:discretization_comparision}. This fact gives us confidence that the discretization errors are well
under control.

\subsection{Lattice measurements}
\begin{figure}
    \includegraphics[width=8.6cm]{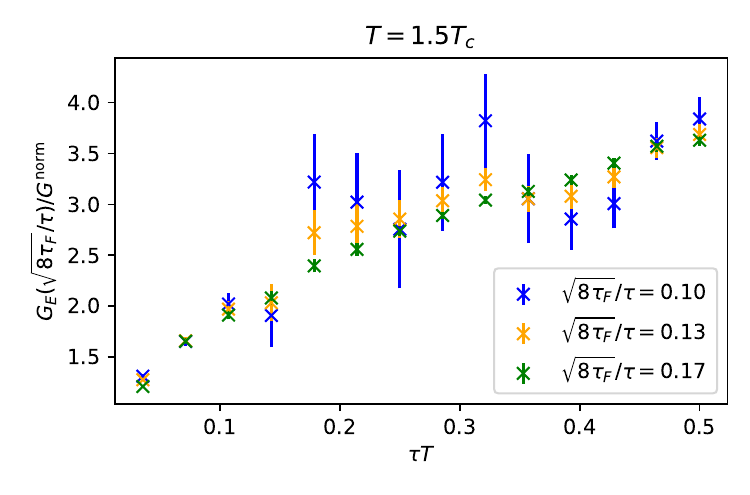}
    \includegraphics[width=8.6cm]{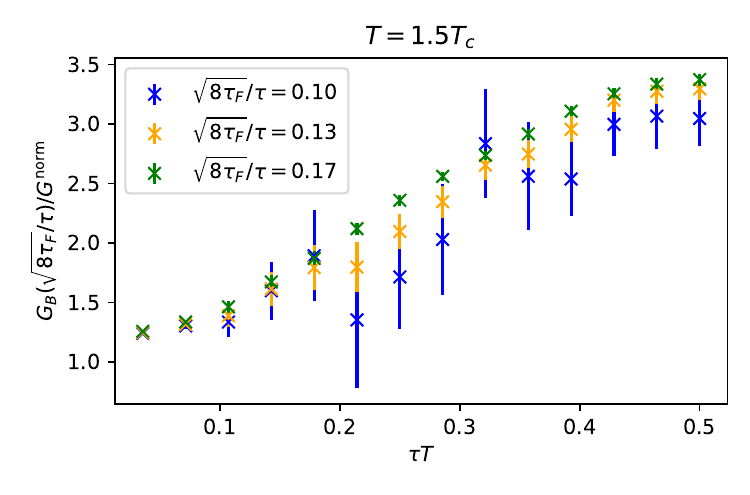}
    \caption{Normalized correlators $\GE$ (top) and $\GB$ (bottom) at fixed flow-time ratios, $\sqrt{8 \tauF}/\tau$ for the $N_t=28$ lattice at $T=1.5T_c$. 
    We see that with increasing ratio, the correlators converge toward a common shape across the whole $\tauT$ range.}
    \label{fig:corr_fixed_ratio_examples}
\end{figure}
In Fig.~\ref{fig:corr_fixed_ratio_examples}, we present both electric and magnetic correlators of the raw lattice data, normalized with Eq.~\eqref{eq:gepert} 
and tree-level improvement at different flow times for a single representative lattice size $\Nt=28$.
We observe the statistical errors decreasing as the ratio $\sfrac{\sqrt{8\tauf}}{\tau}$ increases, and that for $\sfrac{\sqrt{8\tauf}}{\tau}>0.1$ the curves at different flow times
seem to converge toward a common shape. This shape seems to be shared between both $\GE$ and $\GB$.

\begin{figure}
    \includegraphics[width=8.6cm]{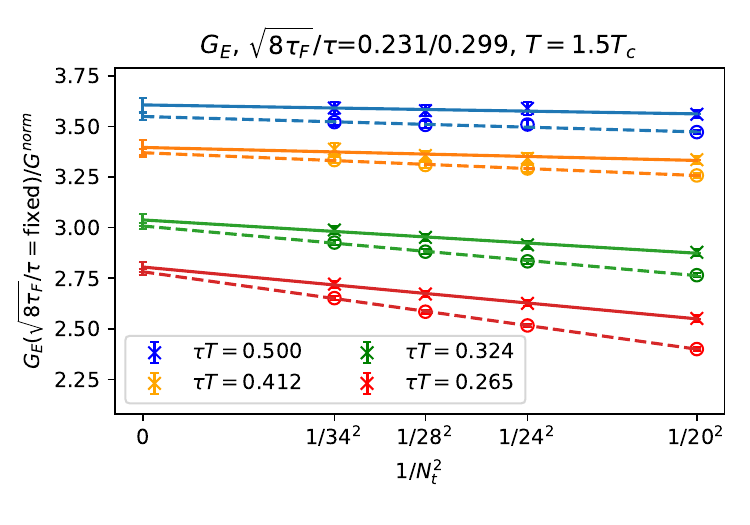}
    \includegraphics[width=8.6cm]{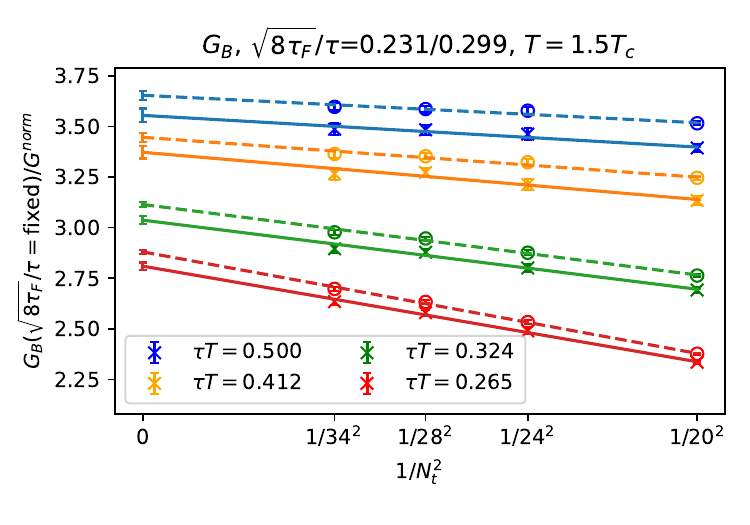}
    \caption{Examples of continuum extrapolations at fixed $\sqrt{8\tauf}/\tau$ for the chromoelectric (top) and chromomagnetic (bottom) correlators at $T=1.5T_c$. The dashed lines and circles indicate the limit taken at the lower edge of the flow-time ratio of interest $\sfrac{\sqrt{8\tauf}}{\tau}=0.231$ while solid lines and asterisks have a higher ratio of $\sfrac{\sqrt{8\tauf}}{\tau}=0.299$. The different $\tauT$ values are shown in different colors.
    }
    \label{fig:cont_limit_example_limit}
\end{figure}
Next, we perform the continuum extrapolations of both correlators. 
First, we interpolate the data for each lattice in $\tauT$ at a fixed flow-time ratio with cubic spline interpolations. 
Since the correlators $\GEB$ are symmetric around the point $\tauT=0.5$, we set the first derivative of the splines equal to zero at $\tauT=0.5$.
We perform a linear extrapolation in $1/\Nt^2 = (aT)^2$ of the correlators at the fixed interpolated $\tauT$, and fixed flow-time ratio positions, using lattices $\Nt = 20,~24,~28,~\mathrm{and}~34$ for large separations
$\tauT>0.25$. For small separations $\tauT<0.25$, we drop the $\Nt=20$ lattice from the extrapolation. 
As an example, we show the continuum extrapolations at different $\tauT$ and $\sfrac{\sqrt{8\tauf}}{\tau}$ 
in Fig.~\ref{fig:cont_limit_example_limit}.
The $\chi^2/df$ of the continuum extrapolation is around 1 or smaller. For small $\tau$ some continuum extrapolations have large
$\chi^2$, indicating that the cutoff effects are too large to
obtain reliable results.
We also perform continuum extrapolations including a $1/\Nt^4$ term for lattices with $\Nt=16$, which corresponds to a $\mathcal{O}(a^4)$ continuum extrapolation. These
continuum extrapolations agree with the ones shown in Fig.~\ref{fig:cont_limit_example_limit} within errors. 
Further details on the continuum extrapolations are discussed in
Appendix~\ref{app:discretization_comparision}.

We present the continuum limits at the edges of the $\sfrac{\sqrt{8\tauf}}{\tau}$ range, within witch we will later take the zero-flow-time limit, 
and see that the continuum values vary less when  $\sfrac{\sqrt{8\tauf}}{\tau}$ is changed than when $\tauT$ is changed. 
Hence, the thermal effects of heavy quark diffusion dominate the shape of these correlators.

\begin{figure}
    \includegraphics[width=8.6cm]{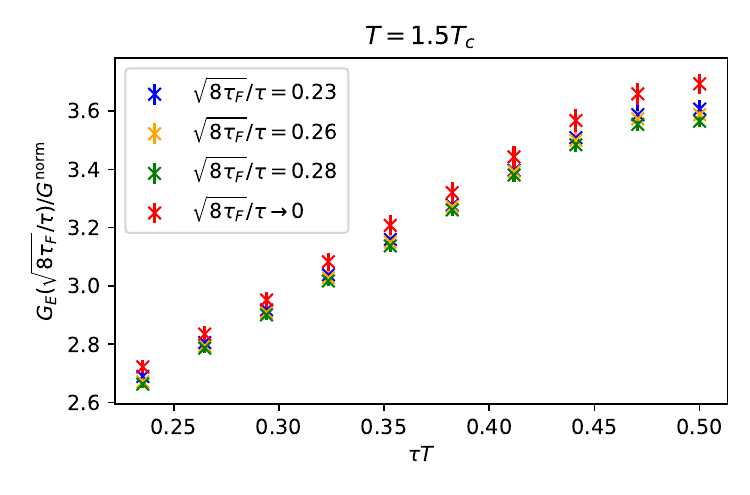}
    \includegraphics[width=8.6cm]{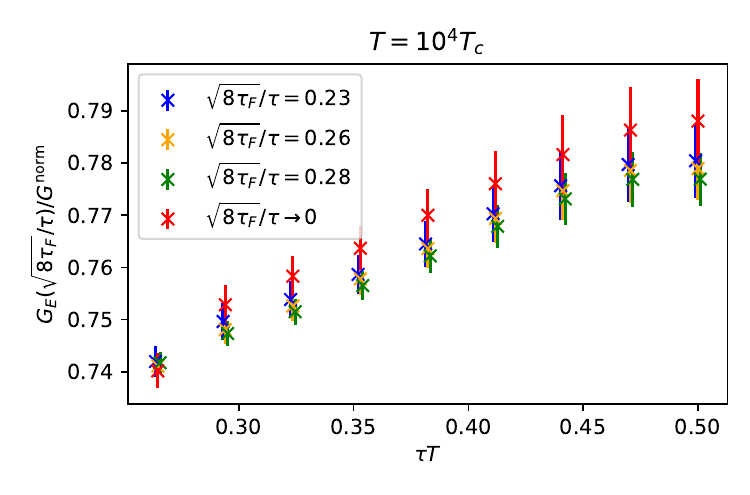}
    \caption{Continuum limit correlators of the chromoelectric correlator $\GE$ 
             at $T=1.5T_c$ (top) and $T=10^4\Tc$ (bottom) for different fixed flow-time ratios and in the zero-flow-time limit.
    }
    \label{fig:cont_limit_example}
\end{figure}
\begin{figure}
    \includegraphics[width=8.6cm]{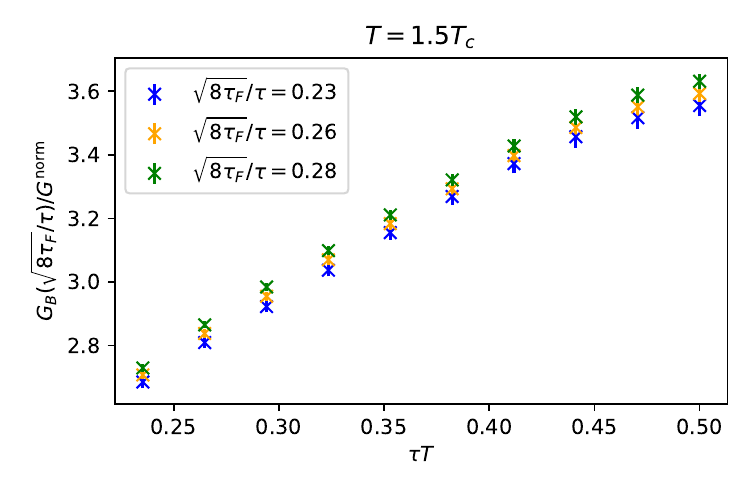}
    \includegraphics[width=8.6cm]{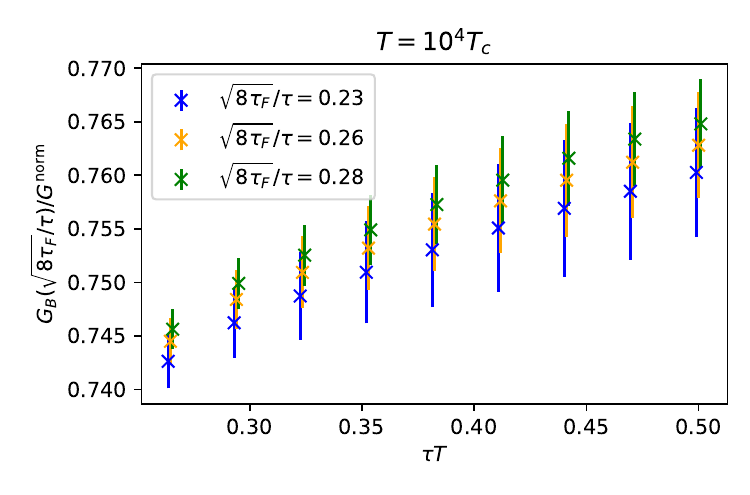}
    \caption{Continuum limit correlators of the chromomagnetic correlator $\GB$ 
             at $T=1.5T_c$ (top) and $T=10^4\Tc$ (bottom) for different fixed flow-time ratios.}
    \label{fig:conth_limit_example}
\end{figure}
\begin{figure}
    \includegraphics[width=8.6cm]{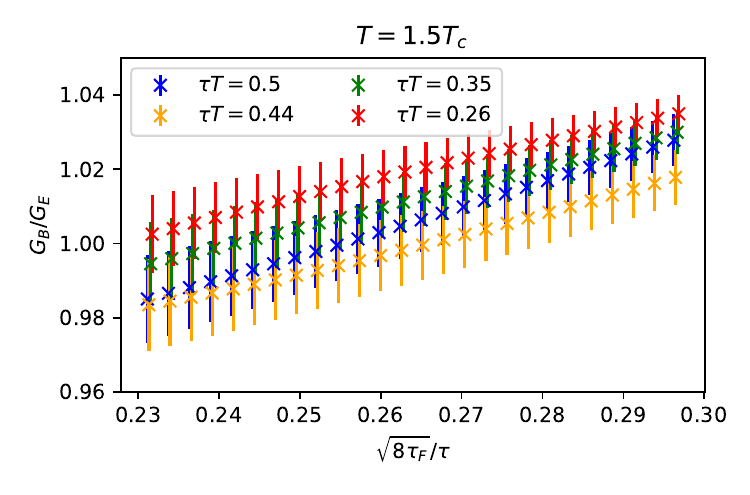}
    \includegraphics[width=8.6cm]{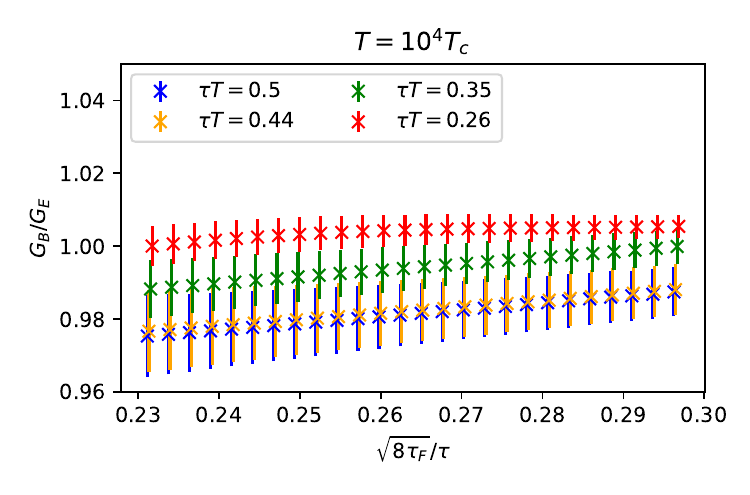}
    \caption{Ratio of the chromomagnetic to the chromoelectric correlator along the fixed flow-time ratio axis for temperatures $T=1.5\Tc$ (top) and $T=10^4\Tc$ (bottom).}
    \label{fig:Gb_Ge_ratios}
\end{figure}
Figures~\ref{fig:cont_limit_example} and~\ref{fig:conth_limit_example} show the final continuum limits of $\GE$ and $\GB$, respectively, for both measured temperatures as function of $\tauT$. 
Similarly to what we observed in Fig.~\ref{fig:corr_fixed_ratio_examples}, both correlators exhibit similar behavior at fixed temperatures according to the shape and order of magnitude. 
As mentioned above for the chromomagnetic correlator, we also perform calculations
using clover discretizations and verify that the same continuum limit is obtained
for this discretization.
This is shown in Appendix~\ref{app:discretization_comparision}.

To further inspect this similarity, in Fig.~\ref{fig:Gb_Ge_ratios} we plot the ratio $\GE/\GB$ along the fixed $\sfrac{\sqrt{8\tauf}}{\tau}$ axis, 
and observe a near-constant behavior toward large separations $\tauT$.
From here, we can already deduce that the contribution to the heavy quark diffusion coefficient from
the chromomagnetic correlator $\GB$ is only going to differ from the contribution of the chromoelectric correlator $\GE$ by less than 5\%. 
In Fig.~\ref{fig:cont_limit_example}, we also show the zero-flow-time limit for the chromoelectric correlator $\GE$, which will be discussed further in the next section.

\section{Measuring the diffusion coefficient on the lattice}\label{sec:GEkappa}
\subsection{Modeling of the spectral function}\label{subsec:modlspec}
We now turn to extracting $\kappaE$ and $\kappaB$ from $\GE$ and $\GB$, respectively, using Eqs.~\eqref{eq:gefromrho}~and~\eqref{eq:kappalimd}.
Our strategy for modeling the spectral function closely follows the approach laid out in our previous
work \cite{Brambilla:2020siz}. This approach uses the perturbative information on the spectral
function at large $\omega$, where this information is expected to be reliable.
For both correlators, the spectral function $\rhoEB$ is known at the NLO level~\cite{Burnier:2010rp,Banerjee:2022uge}. 
We chose to model the spectral function such that in the UV regime at zero flow time it follows the $T=0$ part of the NLO spectral function; 
however, we chose the scale so that the NLO part vanishes, leaving us with only the LO part~\cite{Caron-Huot:2009ncn}:
\be\label{eq:lorho}
\rhoEB^\mathrm{LO}(\omega,T) = \frac{g^2(\mu_\omega^\mathrm{opt})\CF\omega^3}{6\pi}\,.
\ee
The coupling has been evaluated at the five-loop\footnote{As noted in our preceding multilevel study~\cite{Brambilla:2020siz}, 
the results would stay the same even if two-loop running was used.} level in the $\MSb$ scheme at the scale $\mu_\omega^\mathrm{opt}$,
which for $\rhoE$ reads~\cite{Burnier:2010rp},
\be\label{muom1}
\ln(\mu_\omega)=\ln(2\omega) +
\frac{(24\pi^2-149)}{66}\,.
\ee
For the electric spectral function $\rhoE$, we further change to the NLO EQCD scale~\cite{Kajantie:1997tt} 
\be\label{muom2}
\ln(\mu_\omega)= \ln(4\pi T) - \gamma_\mathrm{E} -\frac{1}{22} 
\ee
when $\omega\approx T$ or smaller.
As we discuss below, the LO or NLO result for the UV part of the spectral function
is not accurate, and we hence multiply it by a normalization factor $C_n$
to take into account higher-order corrections, i.e., we perform the replacement
$\rhoEB^\mathrm{LO} \rightarrow C_n \rhoEB^\mathrm{LO}$. A similar normalization
constant was used in the analysis of Refs.~\cite{Francis:2015daa,Altenkort:2020fgs}.
The determination of $C_n$ is discussed at the end of this subsection.
For $\rhoE$, we do not model the flow-time dependence of the spectral function, as one is able to take the zero-flow-time limit before the spectral function inversion.

For the magnetic spectral function $\rhoB$, the situation is more complicated due to the required renormalization~\cite{Bouttefeux:2020ycy,Laine:2021uzs}.
In order to study the chromomagnetic correlator $\GB$ at zero flow time,
we use the relation between the UV part of $G_B$ at nonzero
flow time and the corresponding renormalized correlator in the $\MSb$ scheme:
\ba
\GB^\mathrm{flow, UV}(\tau,\tauF)=&(1+\gamma_0 g^2 \ln(\mu \sqrt{8 \tauF}))^2 \times \nonumber \\ &Z_\mathrm{flow} \GB^{\MSb,\mathrm{UV}}(\tau,\mu) 
+h_0 \cdot(\tauF/\tau)\,,\label{eq:gbgfmsb}
\ea
where $h_0$ is a constant and $\gamma_0=3/(8\pi^2)$ is the anomalous dimension of the chromomagnetic field~\cite{Banerjee:2022uge}. 
In principle, the renormalization constant $Z_\mathrm{flow}$ can be calculated in perturbation theory;
however, in practice we know from our previous calculation~\cite{Brambilla:2020siz} that the NLO perturbative results are  not reliable enough to fully describe the lattice data. 
Hence, $Z_\mathrm{flow}$ is fixed by comparing the perturbative result to the lattice result on $\GB$.
Using the NLO result from Ref.~\cite{Banerjee:2022uge} and neglecting the distortions due to finite flow
time (i.e., setting $h_0$ to zero), Eq.~\eqref{eq:gbgfmsb} gives a flow-time-dependent UV part of the
chromomagnetic spectral density:
\ba
\rhoB^\mathrm{UV}(\omega,\tauF)=&Z_\mathrm{flow} \frac{g^2(\mu) \omega^3}{6 \pi} 
\times \nonumber \\ &(1 + g^2(\mu) (\beta_0-\gamma_0) \ln(\mu^2/(A \omega^2)) 
\nonumber \\ &
+g^2(\mu) \gamma_0 \ln(8 \tauF \mu^2) \,, \label{eq:flowrhob}
\ea
where $\beta_0=11/(16\pi^2)$ is the leading coefficient of the $\beta$ function, and
\be
A = \exp\left[\frac{134}{35} - \frac{8\pi^2}{5} - \ln 4\right]\,.
\ee
As with $\rhoE$, we choose the scale $\mu^\mathrm{opt}$ in such a way that Eqs.~\eqref{eq:flowrhob} and~\eqref{eq:lorho} are equal
up to a constant, $Z_\mathrm{flow}$:
\be\label{muom3}
\mu^\mathrm{opt}=(\sqrt{A} \omega)^{1-\gamma_0/\beta_0} \cdot (8 \tauF)^{-\gamma_0/(2 \beta_0)}\,.
\ee
As in the case of the chromoelectric spectral function, here we make the replacement 
$\rhoB^{\mathrm{UV}} \rightarrow C_n \rhoB^{\mathrm{UV}}$ to take into account higher-order perturbative corrections. The determination of the normalization constant $C_n$ is discussed
below, and now $C_n$ will also contain the unknown normalization factor $Z_\mathrm{flow}$.

The perturbative spectral functions described so far cover the UV regime of our model spectral functions. 
Alone, these UV spectral functions would give $\kappaEB=0$, and hence
an infrared contribution needs to be added leading to finite $\kappaEB$.
We note that while in general the chromomagnetic spectral function depends on the renormalization scheme ($\MSb$, gradient flow, etc.) and scale, its low-frequency limit does not since $\kappaB$ is a physical
quantity. This has been shown explicitly in weak-coupling calculations~\cite{Bouttefeux:2020ycy}. One can work 
with the physical (RG-invariant) chromomagnetic spectral function by scaling out the anomalous dimension, or one can
equally well work with the chromomagnetic correlation function in the gradient flow scheme at some finite,
but sufficiently small, $\tauF$, and extract $\kappaB$.

In order to extract the $\kappaEB$, we then follow the procedure laid out in our preceding study~\cite{Brambilla:2020siz} and model the spectral function with a family of Ans\"atze.
For the large-$\omega$ regime in the UV, we assume the LO perturbative spectral function at $T=0$ as $\rho^\mathrm{UV}$ from Eq.~\eqref{eq:lorho} to hold.
while for small $\omega$ in the IR, the spectral function is given by
\be
\rhoEB^\mathrm{IR}(\omega,T) = \frac{\omega\kappa}{2T}\label{eq:rhoIR}\,.
\ee
We assume that $\rhoEB(\omega,T)=\rho^\mathrm{IR}(\omega,T)$ for $\omega<\omega^\mathrm{IR}$
and $\rhoEB(\omega,T)=\rhoEB^\mathrm{UV}(\omega,T)$ for $\omega>\omega^\mathrm{UV}$,
where $\omega^\mathrm{IR}$ and $\omega^\mathrm{UV}$ are the limiting values of $\omega$ for which we can trust the above behaviors.
In the region $\omega^\mathrm{IR}<\omega<\omega^\mathrm{UV}$, the form of the spectral function is generally not known, 
and this lack of knowledge will generate an uncertainty in the determination of $\kappaEB$. 
Hence, for a given value of $\kappaEB$, we construct the model spectral function that is given by $\rhoEB^\mathrm{UV}$ in $\omega > \omega^\mathrm{UV}$,
$\rhoEB^\mathrm{IR}$ in $\omega<\omega^\mathrm{IR}$, and a variety of forms of $\rhoEB(\omega)$ for the intermediate $\omega^\mathrm{IR} \le \omega \le \omega^\mathrm{UV}$,
such that the total spectral function is continuous.
For the functional forms of the spectral function in the intermediate $\omega$ values, we consider two possible forms based on simple
interpolations between the IR and UV regimes:
\ba
&\rhoEB^\mathrm{line}(\omega,T) = \rhoEB^\mathrm{IR}(\omega,T)\theta(\omega^\mathrm{IR}-\omega) +\nonumber\\ 
&\left[\frac{\rhoEB^\mathrm{IR}(\omega^\mathrm{IR},T)-\rhoEB^\mathrm{UV}(\omega^\mathrm{UV},T)}{\omega^\mathrm{IR}-\omega^\mathrm{UV}}
\left(\omega-\omega^\mathrm{IR}\right)+\rhoEB^\mathrm{IR}(\omega^\mathrm{IR},T)\right] \nonumber \\ &\times
\theta(\omega - \omega^\mathrm{IR})\theta(\omega^\mathrm{UV}-\omega) + \rhoEB^\mathrm{UV}(\omega,T)\theta(\omega-\omega^\mathrm{UV}) 
\label{eq:ans-lin}
\ea
and
\be
\rhoEB^\mathrm{step}(\omega,T) = \rhoEB^\mathrm{IR}(\omega,T) \,\theta(\Lambda-\omega) + 
\rhoEB^\mathrm{UV}(\omega,T)\,\theta(\omega-\Lambda)\,,
\label{eq:ans-step}
\ee
where $\theta(\omega)$ is a step function. The case described in Eq.~\eqref{eq:ans-step} corresponds to $\omega^\mathrm{IR}=\omega^\mathrm{UV}=\Lambda$ 
with the value of $\Lambda$ self-consistently determined, 
i.e., the value of $\Lambda$ is set by requiring the model spectral function to be continuous.
We will refer to these two forms as the line model and the step model, respectively.
In our previous analysis, we determined that the NLO spectral function takes the linear form for ${\omega<0.02T}$, and converges to the UV form at $\omega>2.2T$,
and hence we use the same $\omega^\mathrm{IR}=0.01T$ and $\omega^\mathrm{UV}=2.2T$ as in Ref.~\cite{Brambilla:2020siz} for the line model~\eqref{eq:ans-lin} 
and for both chromomagnetic and chromoelectric spectral functions.
The correlation functions obtained from the model spectral functions through Eq.~\eqref{eq:gefromrho} will be labeled as $\GEB^\mathrm{model}$.

The spectral representation of $\GEB$ given by Eq.~\eqref{eq:gefromrho} also holds at finite lattice spacing ($a\ne 0$) and finite flow time ($\tauF \ne 0$), 
as long as the spectral function $\rhoEB$ is replaced by a lattice equivalent $\rhoEB^\mathrm{lat}(a,\tauf)$.
The spectral function $\rhoEB^\mathrm{lat}(a,\tauF)$ only has support for $\omega<\omega_\mathrm{max}$.
In the case of meson correlators, a similar $\rho^\mathrm{lat}$ has been explicitly constructed in the free case~\cite{Karsch:2003wy}.
In this work, point-like meson sources and sinks are used. One often uses correlation functions of extended meson
operators to improve the signal of the ground state. The spectral function of such extended meson correlators has also been
calculated in the free theory~\cite{Kaczmarek:2003dp}. It was found that for extended meson operators, the support of 
the spectral function shifts toward smaller $\omega$ values~\cite{Kaczmarek:2003dp}, and their shape is modified at large
$\omega$ but not at small $\omega$~\cite{Kaczmarek:2003dp}. The operators obtained from gradient flow
can be viewed as extended operators, and therefore the shape of the corresponding spectral
functions at large $\omega$ will be different compared to the unsmeared case, and the support of the spectral function will
shift toward smaller $\omega$.
However, the small-$\omega$ limit of $\rhoEB^\mathrm{lat}(a,\tauf)$  will not depend on $a$ or $\tauf$ to a good approximation,
because the correlator $\GEB$ is not sensitive to $a$ or $\tauf$, provided that $\tau \gg a$ and $\tau\gg\sqrt{8 \tauf}$.
Therefore, in principle, one can extract $\kappaEB$ even at finite $a$ and $\tauf$.
However, as this is valid only for $\omega<\omega_\mathrm{max}$, the large-$\omega$ part of $\rhoEB^\mathrm{lat}(a,\tauf)$ cannot
be described by the continuum perturbative result. 
We model the UV part of the spectral function $\rhoEB^\mathrm{UV}$ with Eq.~\eqref{eq:lorho}, up to a multiplicative constant. 
The difference between these and the continuum spectral functions is not expected to be large in terms of the correlators $\GEB$ 
at $\tauT>0.25$, which is the relevant $\tau$ range for the determination of $\kappaEB$.

The above statement about the dependence of the spectral function at large $\omega$ on the flow time appears to
contradict the perturbative analysis of Ref.~\cite{Altenkort:2020fgs}. However, we note that in Ref.~\cite{Altenkort:2020fgs} the analytic
continuation was done in terms of the Matsubara frequency, while here we consider continuation in terms of $\tau$: $t \rightarrow -i \tau$.
These two methods of analytic continuations lead to different results, unless the spectral function decays like $1/\omega^2$ for large $\omega$.
Finally, we note that the cutoff effects in $\rho^\mathrm{lat}$ are not limited to the large-$\omega$ region. There are cutoff effects
proportional to $(a T)^2 \sim 1/\Nt^2$ which affect $\rho^\mathrm{lat}$ at all values of $\omega$. However, these are quite small for
the $\Nt>16$ used in our calculations.
\begin{figure}
    \includegraphics[width=8.6cm]{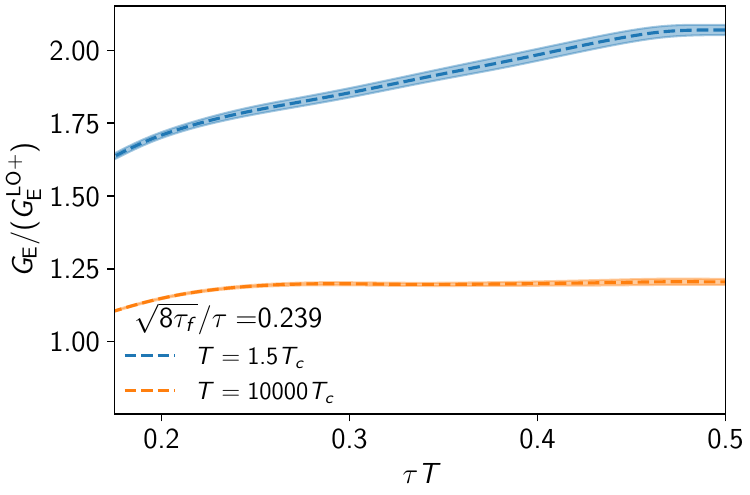}
    \includegraphics[width=8.6cm]{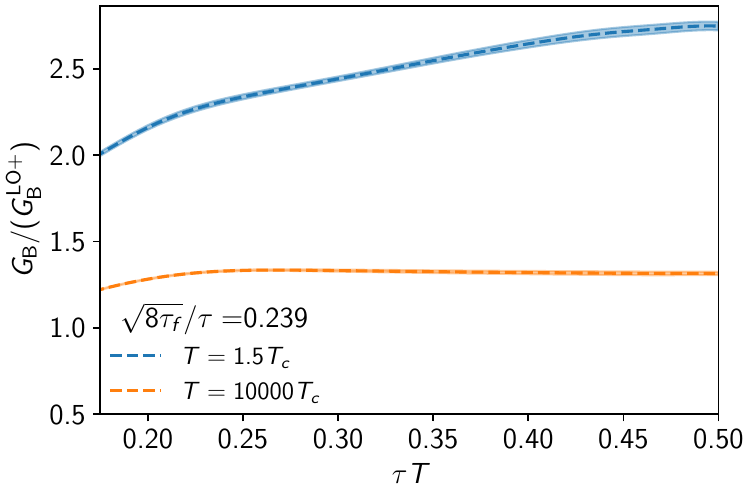}
    \caption{Chromoelectric (top) and the chromomagnetic (bottom) correlators 
             normalized with the LO perturbative result, such that the running coupling is involved.
    }
    \label{fig:gelopnorm}
\end{figure}

So far, we have presented the correlators $\GEB$ normalized with Eq.~\eqref{eq:gepert}, which assumes a constant coupling.
In Fig.~\ref{fig:gelopnorm} we include the running coupling in the analysis, and divide the continuum limit of the correlators
for $\sqrt{8 \tauF}/\tau=0.239$ with Eq.~\eqref{eq:gefromrho}, 
using Eq.~\eqref{eq:lorho}
with the scales~\eqref{muom1} and~\eqref{muom3} for $\rhoE$ and $\rhoB$, respectively.
The corresponding correlators are labeled as $\GEB^\mathrm{LO+}$. We see from Fig.~\ref{fig:gelopnorm} that with this normalization
the $\tau$ dependence of the corresponding ratios is greatly reduced. In particular, at the highest temperature $10^{4} T_c$, 
only very little $\tau$ dependence can be seen for $\tauT \ge 0.25$. The $\tau$ dependence observed for
$\tauT<0.25$ is most likely due to the fact that our continuum extrapolation is not reliable at  such small $\tau$~\cite{Brambilla:2020siz}. Thus, a large part of the $\tau$ dependence of $\GEB$
comes from the running of the coupling constant. On the other hand, the values of the ratios $\GEB/\GEB^\mathrm{LO+}$ differ significantly from one, even at relatively small $\tau$.
A similar trend for $\GE/\GE^\mathrm{LO+}$ was observed in Ref.~\cite{Brambilla:2020siz}.
It was speculated in Ref.~\cite{Brambilla:2020siz} that
the fact that $\GE/\GE^\mathrm{LO+}$ is roughly a constant that is different from one may be due to the one-loop renormalization of the lattice correlator not being reliable.
However, as discussed in Sec.~\ref{subsec:theorybackg}, the one-loop renormalization of the
chromoelectric correlator is quite reliable. This leads us to the conclusion that
the NLO results for the spectral function may not be reliable and an additional normalization
constant, $C_n$, has to be introduced as an extra fit parameter.
The normalization constants $C_n$ are shown
in Appendix~\ref{app:Cn}. For the chromoelectric correlator, the normalization constant $C_n$ is very close to
the one obtained in our study using the multilevel algorithm~\cite{Brambilla:2020siz}. In the case of the chromomagnetic
correlator, the constant $C_n$ also contains the unknown matching between the gradient flow scheme and 
the $\MSb$ scheme, as mentioned before.
We suspect that the fact that the NLO result can describe the lattice correlators
at small $\tau$ only up to a constant $C_n$ is due to 
the presence of the Wilson line
and the Polyakov loop in the definition of the correlators.
These do not contribute at order $g^4$, but will
start contributing at higher orders. It is also known that the weak-coupling result for the Polyakov loop
only works at temperatures $T>5$ GeV~\cite{Bazavov:2016uvm}. At higher orders, the presence of the Wilson
line and the Polyakov loop most likely changes the overall normalization of the correlator, but not its $\tau$ dependence.

\subsection{Flow time dependence of the correlators}
To get rid of distortions due to gradient flow, the lattice results for $\GE$ should be extrapolated  to zero flow time.
The limit to zero flow time has to be taken after the continuum limit to avoid the large discretization effects at small flow times. 
Also, it was argued in Refs.~\cite{Altenkort:2021ntw,Eller:2021qpp} that the inversion of the spectral function via Eq.~\eqref{eq:gefromrho} is mathematically well defined
only in the zero-flow-time limit. As discussed in the previous subsection, it is possible to generalize the spectral
representation in Eq.~\eqref{eq:gefromrho}, for nonzero lattice spacing and flow time, if the corresponding
spectral function only has support for $\omega$ values smaller than some $\omega_\mathrm{max}$.

We also note that in lattice studies of shear viscosity, 
spectral function inversion at finite flow time has given satisfactory results~\cite{Mages:2015rea,Itou:2020azb}. 
Moreover, in recent studies of latent heat, it has been observed that 
the order of the continuum and zero-flow-time limits can be switched 
as long as one is careful to only take the limits in regimes where the functional forms used are justified~\cite{Shirogane:2020muc}.
Therefore, we present our main analysis following the conventional order
continuum limit $\rightarrow$ zero-flow-time limit $\rightarrow$ spectral function inversion for the main analysis,
but we also present an analysis where these steps are taken in a different order. 
\begin{figure}
    \includegraphics[width=8.6cm]{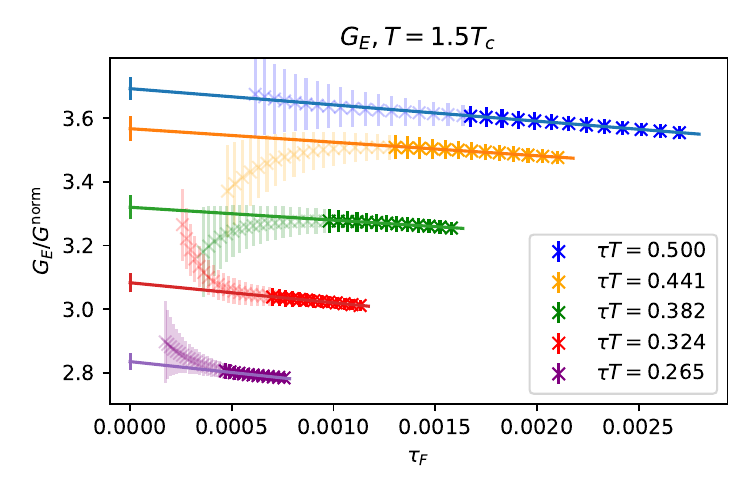}
    \includegraphics[width=8.6cm]{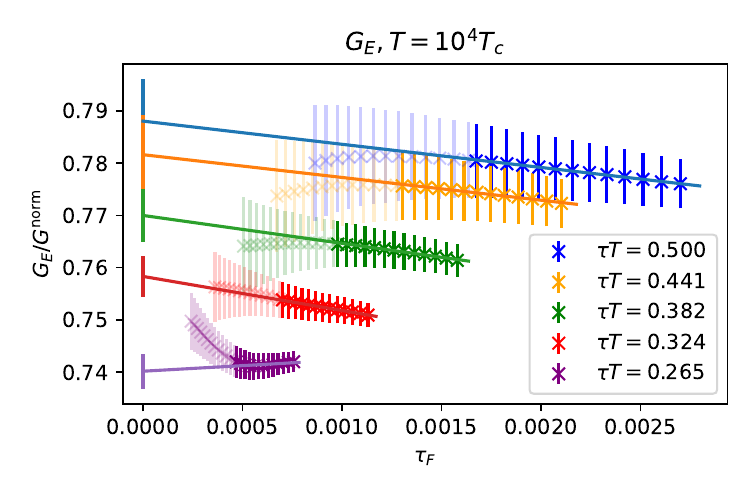}
    \caption{Final results of the continuum limits of the chromoelectric correlator $\GE$ for both temperatures $T=1.5\Tc$ (top) and $T=10^4\Tc$ (bottom). 
             The linear lines indicate the linear zero-flow-time limit. The dimmed symbols correspond to the data not used in the analysis.
             }
    \label{fig:Ge_flowed}
\end{figure}
To perform the extrapolation to the zero-flow-time limit, we use a linear ansatz in $\tauf$.
A linear behavior is expected, as the small-$\tauf$ behavior is just a leading correction to the $\tau$ behavior due to flow.
Moreover, for the chromoelectric correlator $\GE$, the linear behavior has been seen at the NLO level of perturbation theory~\cite{Eller:2021qpp}.
Starting with the chromoelectric correlator $\GE$, we present examples of linear zero-flow-time extrapolations at a few chosen $\tauT$ values in Fig.~\ref{fig:Ge_flowed}.
As expected, we see a clear linear dependence in the range where the extrapolation can be performed.
We observe that the correlator $\GE$ decreases with increasing flow time.
The whole range of $\tauT$ dependence of the zero-flow-time results was already presented in Fig.~\ref{fig:cont_limit_example}.
As one can see from that figure, the flow-time dependence is not very large in the considered flow-time window.
In particular, the shape of the correlator does not change significantly with the flow time and it is very similar
to the shape of the correlator extrapolated to zero flow time. Thus, the determination of $\kappa_E$ 
is not significantly affected by the nonzero flow time.
Therefore, one can also model the spectral function
corresponding to nonzero flow time and determine $\kappaE$. The effects of the small residual distortion of the correlator
due to gradient flow on $\kappaE$ can be taken care of by performing a zero-flow-time extrapolation for the resulting
$\kappaE$. This analysis strategy will be discussed in the next subsection.

\begin{figure}
    \includegraphics[width=8.6cm]{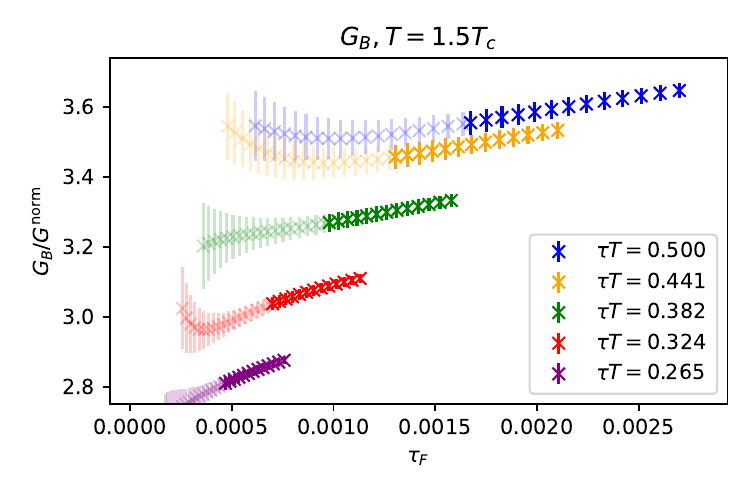}
    \includegraphics[width=8.6cm]{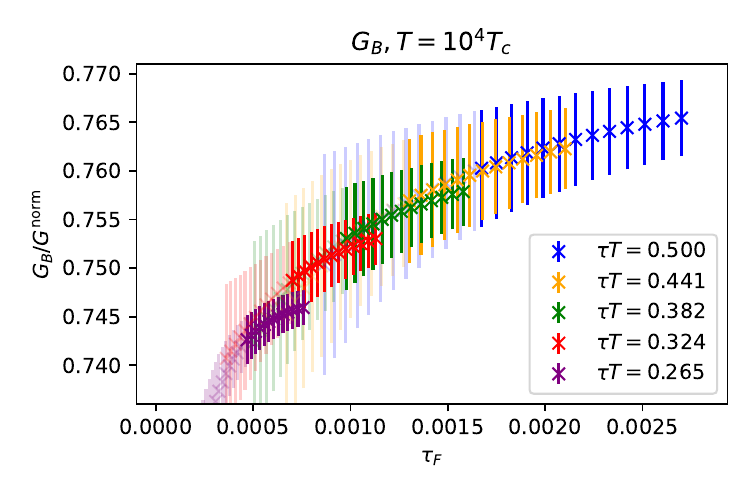}
    \caption{Final results of the continuum limits of the chromomagnetic correlators for both temperatures: $T=1.5\Tc$ (top) and $T=10^4\Tc$ (bottom). The dimmed symbols correspond to the lattice data not used in the determination of $\kappaB$.}
    \label{fig:Gb_flowed}
\end{figure}
\begin{figure}
    \includegraphics[width=8.6cm]{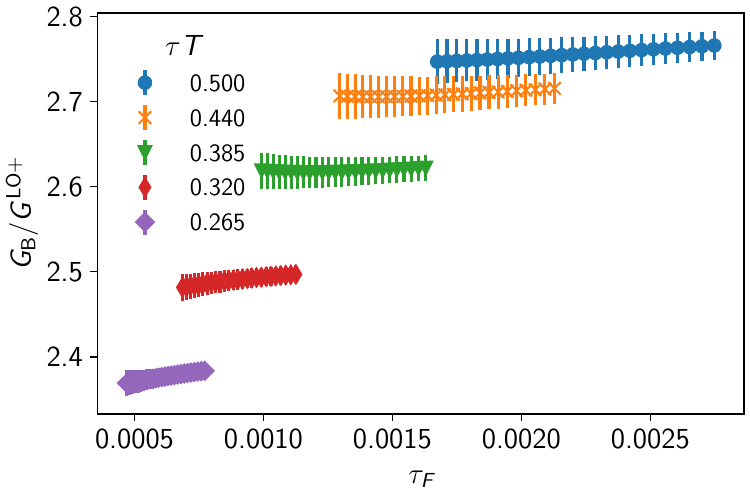}
    \caption{Continuum limit correlators of the chromomagnetic correlator $\GB$ at $T=1.5\Tc$ divided by our default UV model for the spectral function $\rhoB$
      as functions of flow time for a few representative values of $\tau T$. 
 }
    \label{fig:gblopnormtau}
\end{figure}

The flow-time dependence of the chromomagnetic correlator is shown in Fig. \ref{fig:Gb_flowed}, and appears to
be quite different from the flow-time dependence of the chromoelectric correlator. The flow time dependence of
$\GB$ appears to be roughly linear, but its slope has the opposite sign. This difference is expected and probably comes
from the nontrivial renormalization of $\GB$, cf. Eq.~\eqref{eq:gbgfmsb}. This renormalization is taken care
of at leading-order in $\GB^\mathrm{LO+}$. Normalizing the chromomagnetic correlator by $\GB^\mathrm{LO+}$, instead of by 
$G^{\rm norm}$,
largely reduces the flow-time dependence. This is shown in Fig. \ref{fig:gblopnormtau}. In the case of the chromomagnetic
correlator we do not take the zero-flow-time limit, but instead model the spectral function for nonzero flow time using
Eqs.~\eqref{eq:flowrhob},~\eqref{eq:rhoIR},~\eqref{eq:ans-lin},
and~\eqref{eq:ans-step}, and then perform the zero-flow-time
extrapolation of $\kappaB$ obtained from this modeling, as will be described below.

\subsection{Results: $\kappaE$}
The extraction of $\kappaE$ proceeds as follows. 
We take the continuum and zero-flow-time limit data and perform a least-squares fit to Eq.~\eqref{eq:gefromrho} 
with either of the models of the spectral function $\rhoEB(\omega)$.
In addition to having $\kappaEB$ as a fit parameter, 
we also enforce a normalization in the fit by finding a normalization coefficient $C_n(\tau_\mathrm{min}T)$ fit parameter
such that $\sfrac{\GE(\tau_\mathrm{min}T)}{\GE^\mathrm{model}(\tau_\mathrm{min}T)} = 1$. 
To estimate the contributions from the systematic errors, we perform these fits with
different values of $\tauT_\mathrm{min}$, vary the scale $\mu$ of the running coupling by a factor of 2, and perform the fit with either the line~\eqref{eq:ans-lin} or step~\eqref{eq:ans-step} models of the spectral function. To get the final estimate for the heavy quark momentum diffusion coefficient, we then take the full spread of the subset of these fits for which 
the ratio $\sfrac{\GE(\tauT>\tau_\mathrm{min}T)}{\GE^\mathrm{model}(\tauT>\tau_\mathrm{min}T)} = 1$ is within $1.5\sigma$. 

For the $\GE$ data, which was first extrapolated to the continuum limit and then to the zero-flow-time limit, this procedure gives for $\kappaE$
\be
\label{eq:kappaE_1.5Tc}
1.70 \le \frac{\kappaE}{T^3} \le 3.12
\ee
at $T=1.5\Tc$ and 
\be
\label{eq:kappaE_104Tc}
0.02 \le \frac{\kappaE}{T^3} \le 0.16 
\ee
at $T=10^4\Tc$.
The $T=1.5\Tc$ result gives a slightly improved range for $\kappaE$ compared to our previous multilevel study~\cite{Brambilla:2020siz}, 
which had $1.31<\sfrac{\kappaE}{T^3}<3.64$. Although slightly smaller, it is also in agreement 
with the other existing results for this temperature: $2.31<\sfrac{\kappaE}{T^3}<3.70$ from Ref.~\cite{Altenkort:2021ntw}, 
$1.8<\sfrac{\kappaE}{T^3}<3.4$ from Ref.~\cite{Francis:2015daa}, $1.55<\sfrac{\kappaE}{T^3}<3.95$ from Ref.~\cite{Banerjee:2011ra}, and
$1.3<\sfrac{\kappaE}{T^3}<2.8$ from Ref.~\cite{Banerjee:2022uge}.
The $T=10^4\Tc$ result is in agreement with our previous result $0<\sfrac{\kappaE}{T^3}<0.1$~\cite{Brambilla:2020siz}. 
The new result has slightly larger errors due to the gradient flow analysis having more strict fit regimes;
however, we can for the first time observe a nonzero minimum for $\sfrac{\kappaE}{T^3}$ at very large temperature. 
Both of these $\kappaE$ values can be reexpressed as a position-space momentum diffusion coefficient
$D_\mathrm{s}=\sfrac{2T^2}{\kappa}$~\cite{Caron-Huot:2009ncn} as: $0.64<D_\mathrm{s}T<1.17$ for $T=1.5\Tc$ and
$12.5<D_\mathrm{s}T<100$ for $T=10^4\Tc$.

\begin{figure}
    \includegraphics[width=8.6cm]{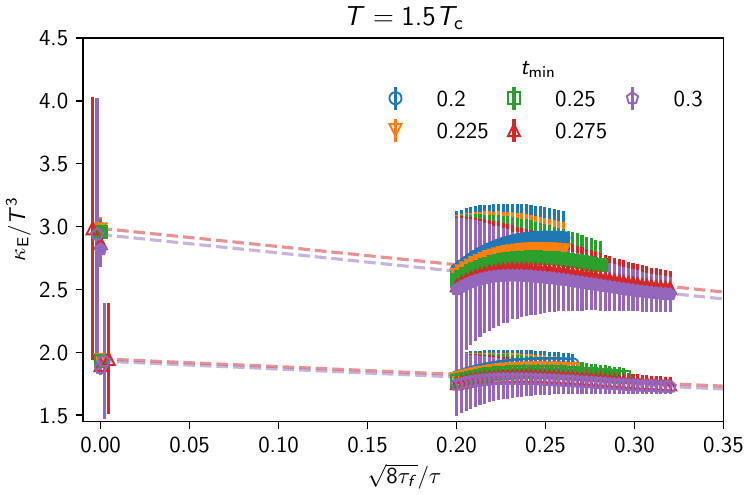}
    \includegraphics[width=8.6cm]{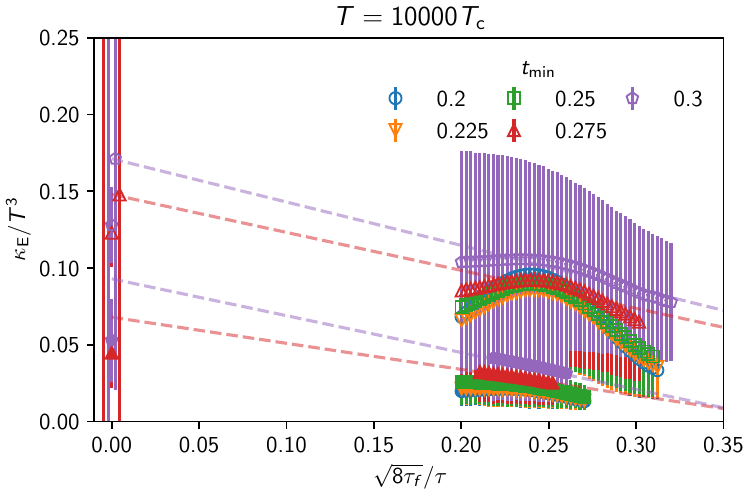}
    \caption{Heavy quark momentum diffusion coefficient $\kappaE/T^3$ at different flow-time ratios $\sqrt{8\tauf}/\tau$ for both temperatures $T=1.5\Tc$ (top) 
             and $T=10^4\Tc$ (bottom). The filled symbols are from the extraction using the line ansatz~\eqref{eq:ans-lin} 
             and the empty symbols are from the step ansatz~\eqref{eq:ans-step} of the spectral function. 
             Different colors depict the different choices for the normalization point $\tauT_\mathrm{min}$. 
             For large $\tauT_\mathrm{min}\ge 0.275$ it is possible to perform linear extrapolation to zero flow time, 
             which is shown as faint dashed lines in color equivalent to the respective $\tauT_\mathrm{min}$.
             }
    \label{fig:Ekappaatfiniteratio}
\end{figure}
\begin{figure}
    \includegraphics[width=8.6cm]{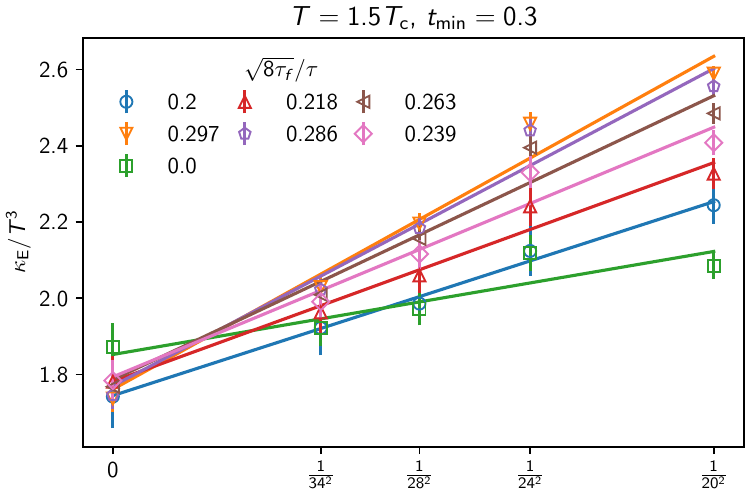}
    \caption{Heavy quark momentum diffusion coefficient $\kappaE/T^3$ extracted at finite lattice spacing for different flow-time ratios $\sqrt{8\tauf}/\tau$ 
             shown with different colors and symbols, for a representative case for $T=1.5\Tc$ and $\tauT_\mathrm{min}=0.3$.
             }
    \label{fig:kappacontlimfiniteratio}
\end{figure}
We now turn to a question of how the result for $\kappaE$ depends on the order of the limits.
First, in Fig.~\ref{fig:Ekappaatfiniteratio} we show the extracted values of $\sfrac{\kappaE}{T^3}$ both at the zero-flow-time limit 
and at a finite flow time for both the line~\eqref{eq:ans-lin} (filled symbols) and the step~\eqref{eq:ans-step} (empty symbols) models for the spectral function $\rho(\omega)$. 
Only points that are within the regime $0.2<\sfrac{\sqrt{8\tauf}}{\tau} < 0.3$ where reasonable zero-flow-time extrapolation can be performed 
and that satisfy the condition $\sfrac{\GE(\tauT>\tauT_\mathrm{min})}{\GE^\mathrm{model}(\tauT>\tauT_\mathrm{min})} = 1$ within $1.5\sigma$ are shown. 
In addition, we show in different colors the different choices of the normalization point $\tauT_\mathrm{min}$. 
We observe that the variation between the models is the dominant source of error and that the variation within the flow time is small in comparison. 
Moreover, with reasonably high $\tauT_\mathrm{min}\ge 0.275$, we have enough data points to perform a linear zero-flow-time extrapolation, 
which we can see agrees closely with the results we get from data that has been extrapolated to zero flow time before spectral function inversion, although with much larger errors. 
If we were to do the $\kappaE$ extraction purely at finite flow time, the full variance due to the different fit forms would give us
$1.5 \le \frac{\kappaE}{T^3} \le 3.2$
for $T=1.5\Tc$ and $0.007 \le \sfrac{\kappaE}{T^3} \le 0.18$ for $T=10^4\Tc$.
Therefore, the variance for a given finite flow time is much larger than the difference between the continuum extrapolated $\kappaE$ extractions. 

Furthermore, we inspect whether it matters that the continuum limit is taken before everything else, as has been done so far.
If we were to instead extract the $\kappaE$ at finite lattice spacing and then take the continuum limit
as a linear extrapolation of the extracted $\kappaE$ values, we would get the result in Fig.~\ref{fig:kappacontlimfiniteratio}. 
We see that the continuum limit of the $\kappaE$ extracted at finite lattice spacing replicates both the zero-flow-time result, 
and the results at finite ratio $\sfrac{\sqrt{8\tauf}}{\tau}$. 
Hence, all results presented above would remain unchanged even if the continuum limit had been taken last, because of the large uncertainties
in $\kappaE$ due to the modeling of the spectral function.

\subsection{Results: $\kappaB$}
We now turn to the chromomagnetic correlator $\GB$ and extraction of the respective $\kappaB$.
Based on the above analysis for $\kappaE$, we can safely assume 
that one can get a very good estimate of the zero-flow-time-extrapolated value even when limiting the analysis to a finite flow time. 
Our analysis strategy here closely follows the case of the chromoelectric correlator.
First, we fix the normalization constant $C_n$ and then vary $\kappaB$ to obtain the best agreement of the lattice correlator
with the model correlator. To demonstrate this point for the step Ansatz, we write
\begin{eqnarray}
&
\displaystyle
\GB^{\rm model}=\frac{\kappaB}{2 T}\int_0^{\Lambda_T} \frac{\dd \omega}{\pi} \omega K(\omega, \tau T)+\nonumber\\[2mm]
&
\displaystyle
C_n(\tauf) \int_{\Lambda_T}^{\infty} \frac{d \omega}{\pi} 
\frac{\CF g^2(\mu_{\omega})}{6 \pi} \omega^3 K(\omega,\tau T)\,,
\end{eqnarray}
where $\Lambda_T \sim T$ is some IR cutoff. We treat $\kappaB$ as a fit parameter, while $C_n(\tau_F)$ is adjusted such that $\GB^{\rm model}$ from
the above equation exactly matches the continuum lattice result for $\GB$ at $\tau=\tau_{\rm min}$.
The values of $C_n$ are shown in Appendix \ref{app:Cn} as a function of $\tauf$.
In Fig. \ref{fig:spf_B} we show the spectral function corresponding to the chromomagnetic
correlators for $\sqrt{8 \tauf}/\tau=0.25$ and $0.3$. As one can see from the figure, the flow-time
dependence of the spectral function is rather mild.
\begin{figure}
\includegraphics[width=8.6cm]{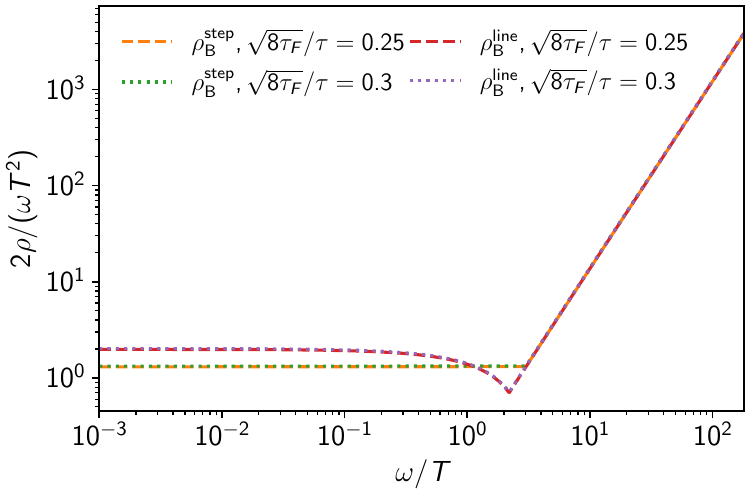}
\caption{Spectral function of chromomagnetic correlators obtained from the two fit forms described in the text at two different representative flow-time ratios.
Only the mean value is shown and the statistical errors are hidden for better visibility.}
\label{fig:spf_B}
\end{figure}
\begin{figure}
    \includegraphics[width=8.6cm]{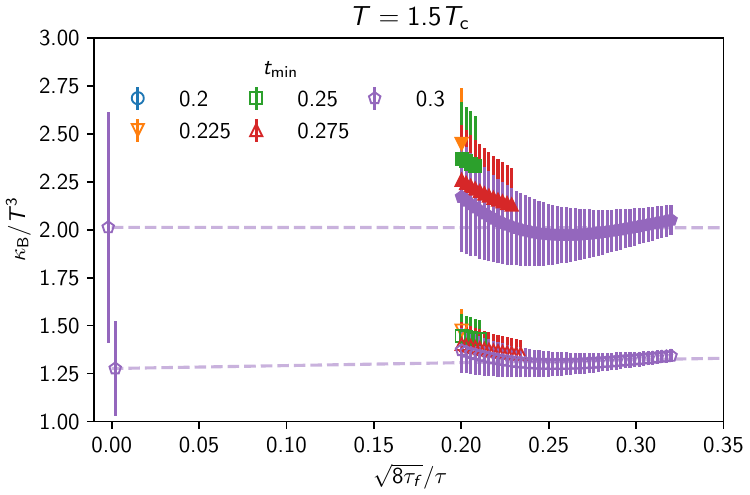}
    \caption{Magnetic heavy quark momentum diffusion coefficient $\kappaB/T^3$ at different flow-time ratios $\sqrt{8\tauf}/\tau$ 
             for  $T=1.5\Tc$. The filled symbols are from extraction using the line ansatz~\eqref{eq:ans-lin} 
             and the empty symbols are from the step ansatz~\eqref{eq:ans-step} of the spectral function. 
             Different colors depict the different choice for the normalization point $\tauT_\mathrm{min}$. 
             The lines and points at $\sqrt{8\tauF}/\tau=0$ depict the zero-flow-time limit taken in the regime $\sqrt{8\tauF}/\tau\ge 0.25$.
    }
    \label{fig:Bkappaatfiniteratio}
\end{figure}

The flow time behavior of the extracted $\sfrac{\kappaB}{T^3}$ is shown in Fig.~\ref{fig:Bkappaatfiniteratio}, 
where again the filled symbols show the extraction using the line ansatz~\eqref{eq:ans-lin},
the empty symbols show the extraction with the step model~\eqref{eq:ans-step}, and different colors depict the different choices of $\tauT_\mathrm{min}$. 
We observe less curvature in the extracted $\kappaB$ values than we saw for $\kappaE$ in Fig.~\ref{fig:Ekappaatfiniteratio}.
If we take the total variation at finite flow time to be the error of $\kappaB$, we get for 
$T=1.5\Tc$ that $1.23< \sfrac{\kappaB}{T^3} < 2.74$. 
We can then proceed to take the zero-flow-time limit in the linear regime $\sqrt{8\tauF}/\tau \ge 0.25$, 
similar to what we learned to work with in the case of $\GE$. In the zero-flow-time limit, we get the final result for $\kappaB$:
\be
\label{eq:kappaB_1.5Tc}
1.03 \le \frac{\kappaB}{T^3} \le 2.61\,.  
\ee
This result is well in agreement with the recent result~\cite{Banerjee:2022uge} of $1.0 \le \kappaB/T^3 \le 2.1$. 
The current data is not accurate enough to determine $\kappaB$ at $T=10^4 T_c$.

\section{Conclusions}\label{sec:conc}
In this paper, we have studied the chromoelectric and chromomagnetic correlators in quenched QCD with the aim
to determine the heavy quark diffusion coefficient, including the subleading correction in the inverse 
quark mass. We used gradient flow for noise reduction and showed how to control the distortions 
due to nonzero flow time in the calculations of the transport coefficients $\kappaE$ and
$\kappaB$. To obtain the heavy quark diffusion coefficient, we used a parametrization
of the spectral functions that relies on the NLO result at large energies, and smoothly matched
to the expected linear behavior at small energies. The effects of the nonzero flow time
can be incorporated into the high energy part of the spectral function. We verified this
in the calculations of $\kappaE$, where we obtained $\kappaE$ from the chromoelectric
correlator extrapolated to zero flow time, as well as by calculating an effective $\kappaE$ from
the chromoelectric correlator at finite flow time and then extrapolating to zero flow time.
Our main results are summarized in Eqs.~\eqref{eq:kappaE_1.5Tc},~\eqref{eq:kappaE_104Tc} and~\eqref{eq:kappaB_1.5Tc}.
Our results for $\kappaE$ agree with the previous determinations~\cite{Kaczmarek:2014jga,Brambilla:2020siz,Altenkort:2021ntw}
within the estimated uncertainties. The value of $\kappaB$ we obtained agrees with the very
recent result obtained using the multilevel algorithm and nonperturbative renormalization based
on the Schr\"odinger functional~\cite{Banerjee:2022uge}. 
We have seen that the dominant uncertainty in the determination of $\kappaE$ and $\kappaB$ comes from the modeling of
the spectral functions at low energies.
Using the lattice results for $\langle \mathbf{v}^2 \rangle$
from Ref.~\cite{Petreczky:2008px} for charm and bottom quarks $\langle \mathbf{v}^2 \rangle_\mathrm{charm} \simeq 0.51$
and $\langle \mathbf{v}^2 \rangle_\mathrm{bottom}\simeq 0.3$ (c.f. Fig. 6 of Ref.~\cite{Petreczky:2008px} where
$v_{th}^2=\langle \mathbf{v}^2 \rangle/3$ was shown), we estimate that the mass-suppressed effect on
the heavy quark diffusion coefficient is 34\% and 20\% for charm and bottom quarks, respectively.

The extraction of the heavy quark diffusion constant strongly relies on using the NLO result for the spectral
function at large energies. It is assumed that the NLO result can describe the $\tau$ dependence of the correlators
up to a multiplicative constant. To test this assertion further, it would be desirable to perform calculations at larger
$N_{\tau}$, so that reliable continuum extrapolations are possible for smaller values of $\tau$. Another way to obtain
more reliable continuum-extrapolated results is to use the Symanzik-improved gauge action. We plan to implement such an improved
analysis in the near future. Finally, once the full one-loop perturbative matching between the $\MSb$ scheme 
and the gradient flow scheme at small flow times becomes available, we will redo our analysis by converting to the $\MSb$ scheme
and taking the zero-flow-time limit.

\begin{acknowledgments}
We thank Antonio Vairo for enriching discussions, and Zeno Kordov for excellent proofreading.
The simulations were carried out on the computing facilities of the Computational Center for Particle and Astrophysics
(C2PAP) in the project \emph{Calculation of finite $T$ QCD correlators} (pr83pu).
This research was funded by the Deutsche Forschungsgemeinschaft (DFG, German Research Foundation) cluster of excellence ``ORIGINS'' (\href{www.origins-cluster.de}{www.origins-cluster.de}) under Germany's Excellence Strategy EXC-2094-390783311.
The lattice QCD calculations have been performed using the publicly available
\href{https://web.physics.utah.edu/~detar/milc/milcv7.html}{MILC code}.
P. P. was supported by the U.S. Department of Energy under 
Contract No. DE-SC0012704.

\end{acknowledgments}

\appendix
\section{Discretization effects and continuum extrapolations}\label{app:discretization_comparision}

\begin{figure*}
    \centering
    \includegraphics[width=17.2cm]{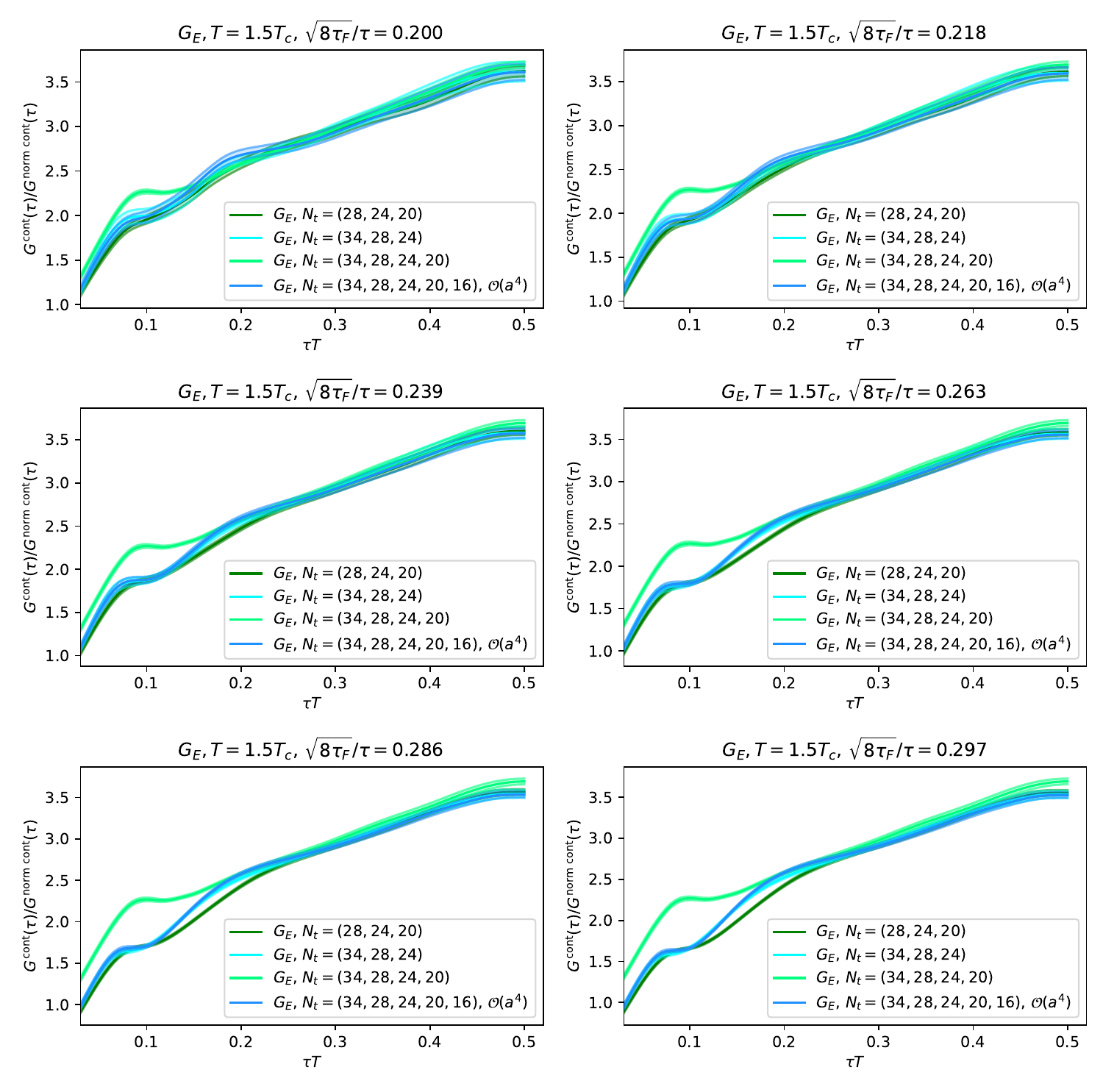}
    \caption{Chromoelectric correlator with different continuum extrapolations for some representative flow times. }
    \label{fig:GE_contdemo}
\end{figure*}

In this appendix, we discuss discretization effects and continuum
extrapolations in more detail. In Fig.~\ref{fig:GE_contdemo} we show 
different continuum extrapolations for the chromoelectric correlators
as a function of $\tau$ for a few representative values of the flow time.
We perform extrapolations assuming a $1/\Nt^2$ form for the discretization
errors and vary the range in $\Nt$, and also include a
$1/\Nt^4$ term in the continuum extrapolations with $\Nt=16$ lattices.
For $\tauT>0.25$ different continuum extrapolations agree well with each
other.
\begin{figure*}
    \centering
    \includegraphics[width=17.2cm]{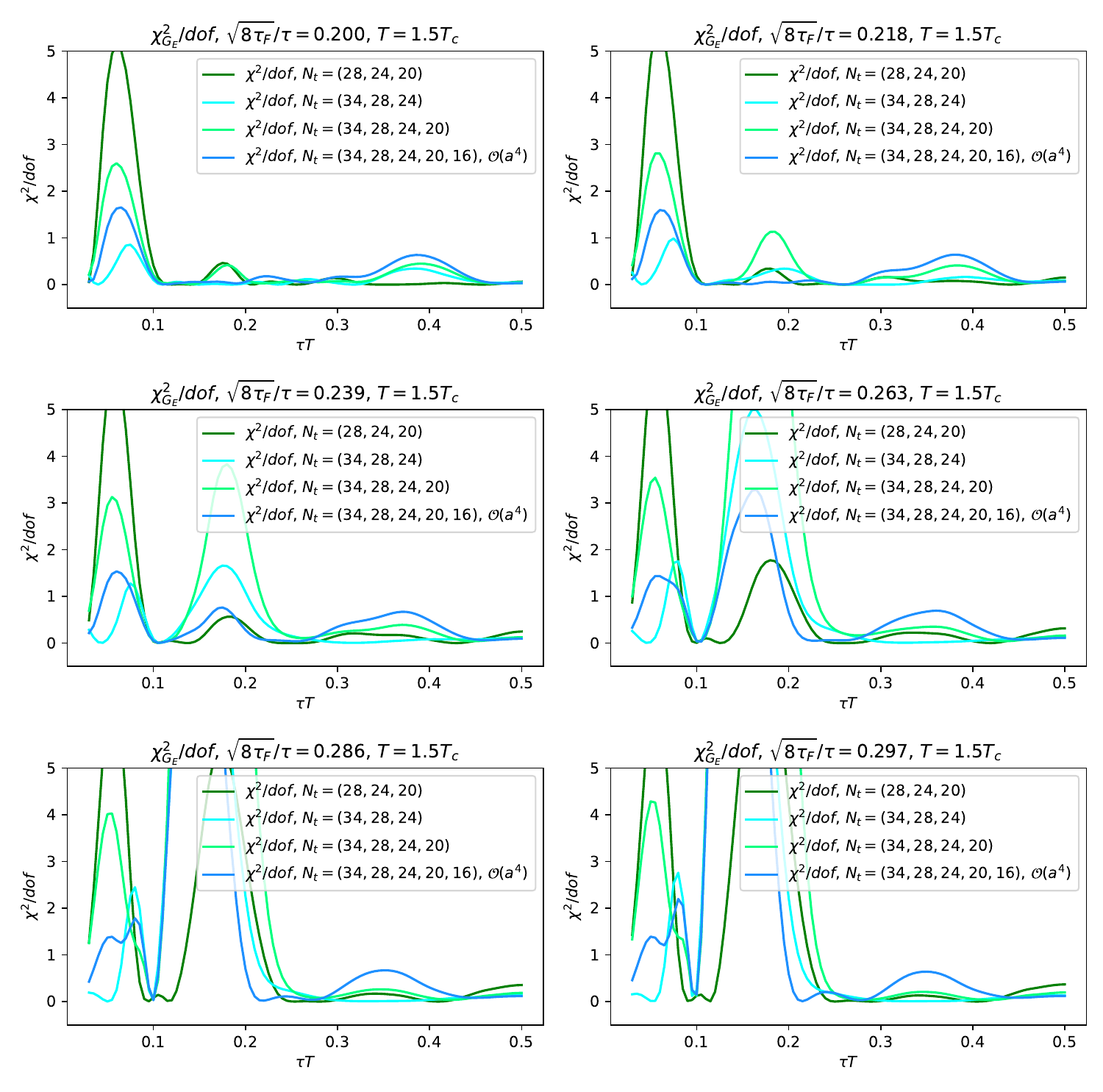}
    \caption{$\chi^2/df$ of different continuum extrapolations of the chromoelectric correlators for some representative flow times. }
    \label{fig:chi2_demo}
\end{figure*}

\begin{figure*}
    \centering
    \includegraphics[width=17.2cm]{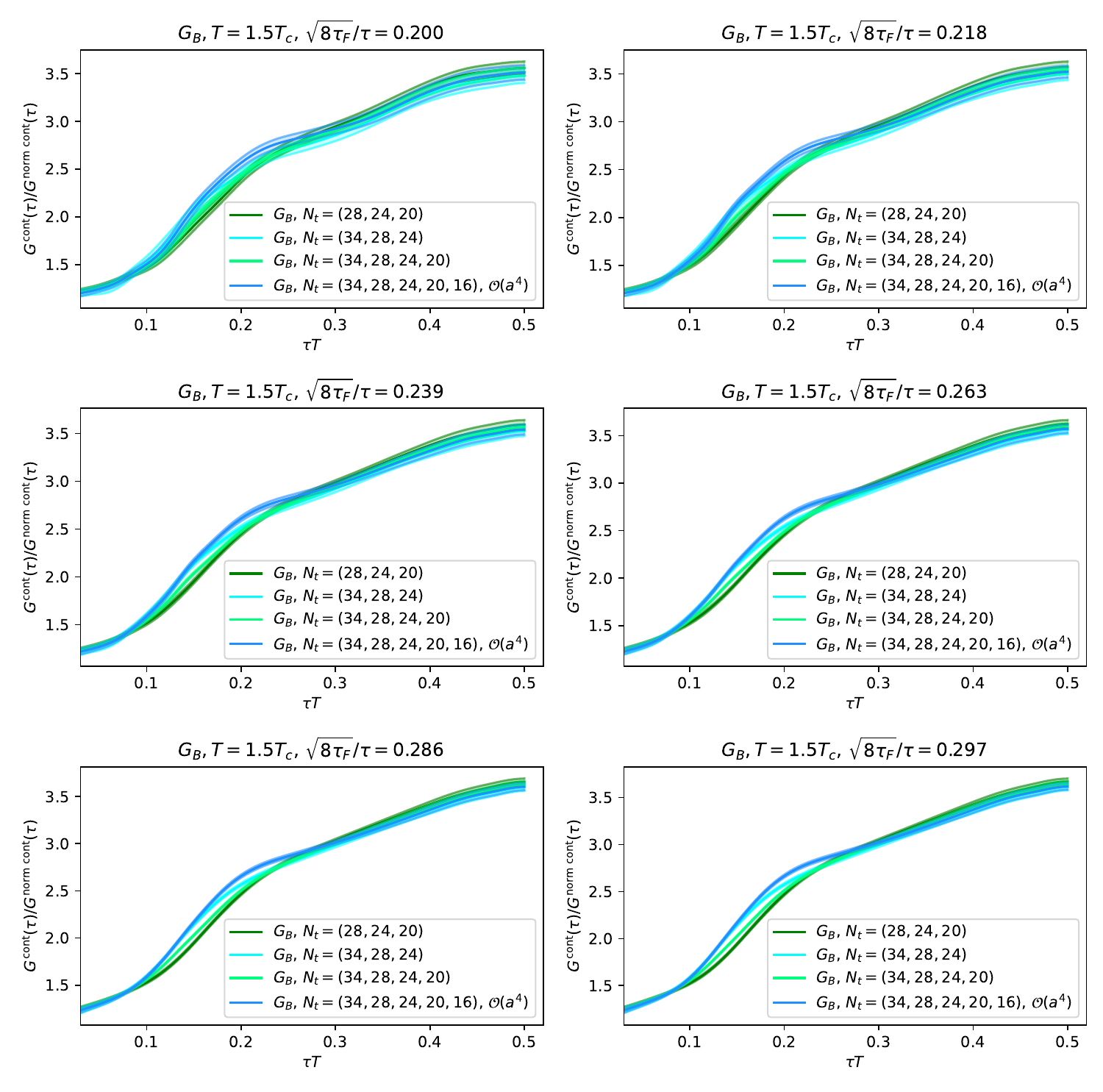}
    \caption{Chromomagnetic correlator with different continuum extrapolations for some representative flow times. }
    \label{fig:GB_contdemo}
\end{figure*}

The $\chi^2/df$ of the continuum extrapolations are
shown in Fig.~\ref{fig:chi2_demo}. For $\tauT>0.25$, different
continuum extrapolations of the chromoelectric correlators agree well, and the $\chi^2/df$ 
of the continuum extrapolation is close to one or smaller. For smaller $\tauT$ the $\chi^2/df$
is large, indicating that the continuum extrapolations are not reliable.
We perform a similar analysis for the chromomagnetic correlators. Some results are shown
in Fig.~\ref{fig:GB_contdemo}
\begin{figure}
    \centering
    \includegraphics[width=8.6cm]{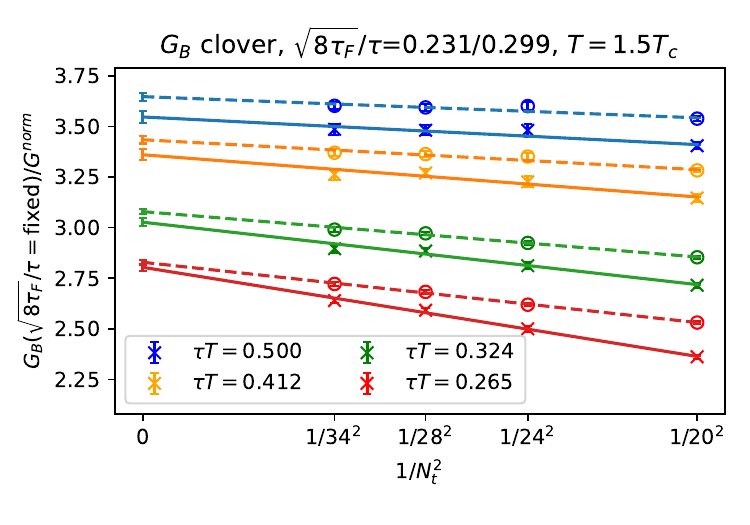}
    \caption{Continuum extrapolation of the chromomagnetic correlator with the clover discretization and the 
    corresponding tree-level improvement~\eqref{eq:clovertreegb}. The results are shown for two different flow times indicated in the label of the figure.}
    \label{fig:GbC_newLOLatt_continuum_limit}
\end{figure}
\begin{figure}
    \centering
    \includegraphics[width=8.6cm]{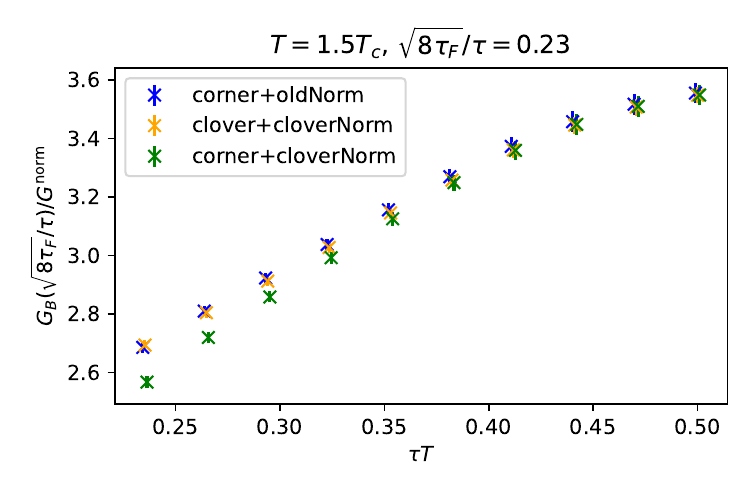}
    \caption{We compare the results of the continuum limits for the following three cases: 
    the corner discretization~\eqref{eq:B_field_loop} normalized with Eq.~\eqref{eq:gepert} (corner+cornerNorm), 
    the clover discretization normalized with Eq.~\eqref{eq:clovertreegb} (clover+cloverNorm),
    the and corner discretization normalized with Eq.~\eqref{eq:clovertreegb} (corner+cloverNorm). }
    \label{fig:Gb_different_norms_compared}
\end{figure}

As discussed in the main text for the chromomagnetic correlators, we use two discretization schemes: the simplest one
given by Eq.~\eqref{eq:B_field_loop}, which can be labeled as the corner discretization, 
and the clover discretization which was also used in Ref.~\cite{Banerjee:2022uge} (c.f. Eqs. (2.2)-(2.4) herein).
These two discretizations must agree in the continuum, but could lead to quite different result at nonzero lattice spacings. 
As a result, the tree-level improvements for these two discretization schemes are also different.
The leading-order result for the clover discretization without the $g^2$ and $C_F$ factors has the form
\begin{align}
    G_\mathrm{norm}^\mathrm{Latt}(\tau)= \frac{1}{3a^4}\int_{-\pi}^\pi &\frac{\mathrm{d}^3\mathbf{q}}{(2\pi)^3}\frac{e^{\Bar{q}N_t(1-\tauT)}+e^{\Bar{q}N_t\tauT}}{e^{\Bar{q}N_t}-1}\nonumber\\
    &\times\frac{\Tilde{q}-\frac{(\Tilde{q}^2)^2+\Tilde{q}^4}{8}+\frac{\Tilde{q}^2\Tilde{q}^4-\Tilde{q}^6}{32}}{\sinh \Bar{q}}, \label{eq:clovertreegb}
\end{align}
where $\Bar{q}$ and $\Tilde{q}$ are given by Eqs.~\eqref{eq:barq} and~\eqref{eq:tildeq} respectively.
We use this to implement the tree level improvement for the clover discretization scheme.
In Fig.~\ref{fig:GbC_newLOLatt_continuum_limit} we show the continuum limit of the flowed chromomagnetic correlator with the clover discretization 
and the tree-level improvement~\eqref{eq:clovertreegb}. We show results for two different flow times, for which the expected $1/\Nt^2$ behavior can be clearly
seen in the lattice data in both cases. We also compare the continuum extrapolated results obtained with the corner and clover discretizations, 
and the corresponding tree-level improvements, in Fig.~\ref{fig:Gb_different_norms_compared}. As one
can see from the figures, the continuum results obtained with the two discretization schemes are in excellent agreement.
The tree-level improvement reduces the discretization effects and therefore, aids  robust continuum extrapolations. However, as discussed
in Ref.~\cite{Brambilla:2020siz} it is not necessary if the lattice spacing is sufficiently small or, equivalently, if $\Nt$ is large enough.
Small lattice spacings are needed for reliable continuum extrapolation at small $\tauT$. If $\tauT$ is not very small, the continuum extrapolation
can be performed without tree-level improvement~\cite{Brambilla:2020siz}. To check to what extent our conclusions on the continuum result
of the chromomagnetic correlator depend on the 
tree-level improvement, we perform continuum extrapolations of the chromomagnetic correlator with the corner discretization scheme but using
the "wrong" tree-level improvement, namely, the tree-level improvement for the clover discretization. The corresponding continuum results are also 
shown in Fig.~\ref{fig:Gb_different_norms_compared} and labeled as "corner+cloverNorm". For $\tauT<0.35$ we see small, but statistically significant,
differences compared to the continuum results obtained with proper tree-level improvement, but for larger values of $\tauT$ the tree-level improvement is not essential for reliable continuum extrapolations.

\section{Normalization parameter}
\label{app:Cn}
\begin{figure}
    \includegraphics[width=8.6cm]{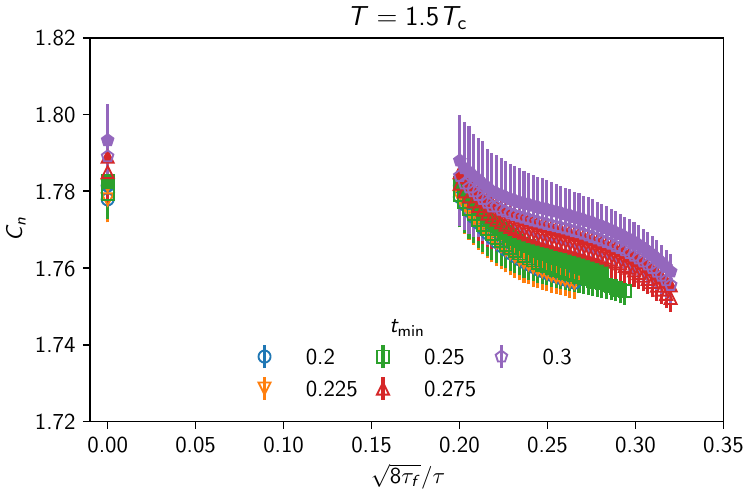}
    \includegraphics[width=8.6cm]{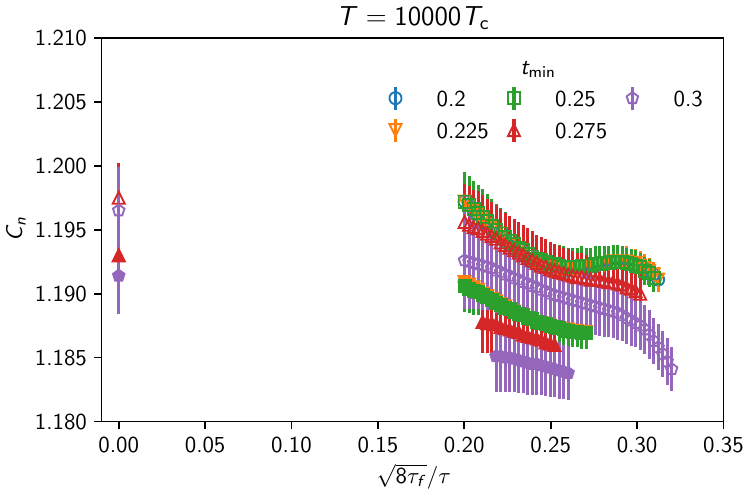}
    \caption{Normalization coefficient $C_n$ for the chromoelectric correlator $\GE$ at different flow-time ratios $\sfrac{\sqrt{8\tauf}}{\tau}$ for both temperatures $T=1.5\Tc$ (top) and $T=10^4\Tc$ (bottom). The filled symbols are from extraction using the line ansatz~\eqref{eq:ans-lin} and the empty symbols are from the step ansatz~\eqref{eq:ans-step} of the spectral function. Different colors depict the different choice for the normalization point $\tauT_\mathrm{min}$. }
    \label{fig:Cnfiniteratio}
\end{figure}
\begin{figure}
    \includegraphics[width=8.6cm]{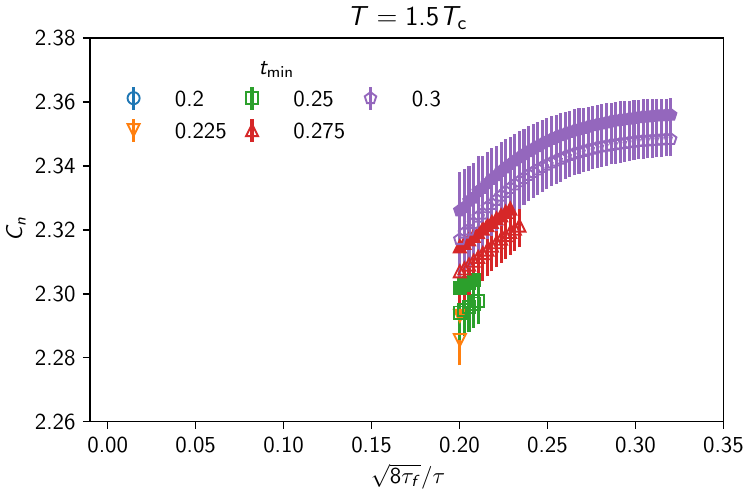}
    \caption{Normalization coefficient $C_n$ for the chromomagnetic correlator $\GB$ at different flow-time ratios $\sfrac{\sqrt{8\tauf}}{\tau}$ for the temperature $T=1.5\Tc$. The filled symbols are from extraction using the line ansatz~\eqref{eq:ans-lin} and the empty symbols are from the step ansatz~\eqref{eq:ans-step} of the spectral function. Different colors depict the different choice for the normalization point $\tauT_\mathrm{min}$. 
    }
    \label{ffig:Cnfiniteratio_kappaB}
\end{figure}
For completeness, we also show in Figs.~\ref{fig:Cnfiniteratio} and~\ref{ffig:Cnfiniteratio_kappaB} the normalization coefficient $C_n$ for both $\GE$ and $\GB$ respectively.
We observe that $C_n$ has a very mild dependence on the flow time. 
This can be used as an indication that modeling $\rhoEB^\mathrm{lat}$ with the running coupling version of the leading-order $\rhoEB$ is reasonably well motivated.
The $C_n$ values for the chromoelectric correlator are well in agreement with the ones we reported in our preceding study~\cite{Brambilla:2020siz}: $\sim1.73$ for $T=1.5\Tc$ and $\sim1.2$ for $T=10^4\Tc$. The $C_n$ for the chromomagnetic field is slightly larger than the respective factor for $\GE$.

%
%

\FloatBarrier
\bibliography{kappa}{}

\end{document}